\documentstyle[preprint,epsf,eqsecnum,aps]{revtex}

\newcommand{\postscript}[2]
 {\setlength{\epsfxsize}{#2\hsize}
  \centerline{\epsfbox{#1}}}

\begin{document}

\tightenlines

\draft
\preprint{FERMILAB--PUB--99/019--E}

\title{ \boldmath Measurement of $B^0 - \bar{B}^0$
        Flavor Oscillations \\
        Using Jet-Charge and Lepton Flavor Tagging \\ 
        in $p\bar{p}$ Collisions at $\sqrt{s}=1.8$ TeV}

\author{ The CDF Collaboration }

\date{\today}

\maketitle

\narrowtext

\def\bd{B^0}
\def\antibd{\bar{B}^0}
\def\bs{B^0_s}
\def\antibs{\bar{B}^0_s}
\def\bplus{B^+}
\def\bminus{B^-}
\def\bbbar{b\bar{b}}
\def\ccbar{c\bar{c}}
\def\cp{CP}
\def\pb{pb$^{-1}$}
\def\ppbar{p\bar{p}}
\def\vtd{V_{td}}
\def\vts{V_{ts}}
\def\vtb{V_{tb}}
\def\dmd{\Delta m_d}
\def\dms{\Delta m_s}
\def\z0{Z^0}
\def\pt{p_T}
\def\et{E_T}
\def\dzero{d_0}
\def\gevc{GeV/$c$}
\def\ptb{p_T^{B}}
\def\ptbprime{p_T^{B\prime}}
\def\ptrel{p_T^{\rm rel}}
\def\mcl{m^{\rm cl}}
\def\ptcl{p_T^{\rm cl}}

\font\eightit=cmti8
\def\r#1{\ignorespaces $^{#1}$}
\hfilneg
\begin{sloppypar}
\noindent
F.~Abe,\r {17} H.~Akimoto,\r {39}
A.~Akopian,\r {31} M.~G.~Albrow,\r 7 S.~R.~Amendolia,\r {27} 
D.~Amidei,\r {20} J.~Antos,\r {33} S.~Aota,\r {37}
G.~Apollinari,\r {31} T.~Arisawa,\r {39}
% ...
\end{sloppypar}
\vskip .026in
\begin{center}
(CDF Collaboration)
\end{center}

\vskip .026in
\begin{center}
\r 1  {\eightit Argonne National Laboratory, Argonne, Illinois 60439} \\
\r 2  {\eightit Istituto Nazionale di Fisica Nucleare, University of Bologna,
I-40127 Bologna, Italy} \\
\r 3  {\eightit Brandeis University, Waltham, Massachusetts 02254} \\
\r 4  {\eightit University of California at Los Angeles, Los 
Angeles, California  90024} \\  
\r 5  {\eightit University of Chicago, Chicago, Illinois 60637} \\
\r 6  {\eightit Duke University, Durham, North Carolina  27708} \\
\r 7  {\eightit Fermi National Accelerator Laboratory, Batavia, Illinois 
60510} \\
\r 8  {\eightit University of Florida, Gainesville, Florida  32611} \\
\r 9  {\eightit Laboratori Nazionali di Frascati, Istituto Nazionale di Fisica
               Nucleare, I-00044 Frascati, Italy} \\
\r {10} {\eightit University of Geneva, CH-1211 Geneva 4, Switzerland} \\
\r {11} {\eightit Harvard University, Cambridge, Massachusetts 02138} \\
\r {12} {\eightit Hiroshima University, Higashi-Hiroshima 724, Japan} \\
\r {13} {\eightit University of Illinois, Urbana, Illinois 61801} \\
\r {14} {\eightit Institute of Particle Physics, McGill University, Montreal 
H3A 2T8, and University of Toronto,\\ Toronto M5S 1A7, Canada} \\
\r {15} {\eightit The Johns Hopkins University, Baltimore, Maryland 21218} \\
\r {16} {\eightit Institut f\"{u}r Experimentelle Kernphysik, 
Universit\"{a}t Karlsruhe, 76128 Karlsruhe, Germany} \\
\r {17} {\eightit National Laboratory for High Energy Physics (KEK), Tsukuba, 
Ibaraki 305, Japan} \\
\r {18} {\eightit Ernest Orlando Lawrence Berkeley National Laboratory, 
Berkeley, California 94720} \\
\r {19} {\eightit Massachusetts Institute of Technology, Cambridge,
Massachusetts  02139} \\   
\r {20} {\eightit University of Michigan, Ann Arbor, Michigan 48109} \\
\r {21} {\eightit Michigan State University, East Lansing, Michigan  48824} \\
\r {22} {\eightit University of New Mexico, Albuquerque, New Mexico 87131} \\
\r {23} {\eightit The Ohio State University, Columbus, Ohio  43210} \\
\r {24} {\eightit Osaka City University, Osaka 588, Japan} \\
\r {25} {\eightit Universita di Padova, Istituto Nazionale di Fisica 
          Nucleare, Sezione di Padova, I-35131 Padova, Italy} \\
\r {26} {\eightit University of Pennsylvania, Philadelphia, 
        Pennsylvania 19104} \\   
\r {27} {\eightit Istituto Nazionale di Fisica Nucleare, University and Scuola
               Normale Superiore of Pisa, I-56100 Pisa, Italy} \\
\r {28} {\eightit University of Pittsburgh, Pittsburgh, Pennsylvania 15260} \\
\r {29} {\eightit Purdue University, West Lafayette, Indiana 47907} \\
\r {30} {\eightit University of Rochester, Rochester, New York 14627} \\
\r {31} {\eightit Rockefeller University, New York, New York 10021} \\
\r {32} {\eightit Rutgers University, Piscataway, New Jersey 08855} \\
\r {33} {\eightit Academia Sinica, Taipei, Taiwan 11530, Republic of China} \\
\r {34} {\eightit Texas A\&M University, College Station, Texas 77843} \\
\r {35} {\eightit Texas Tech University, Lubbock, Texas 79409} \\
\r {36} {\eightit Istituto Nazionale di Fisica Nucleare, University of Trieste/
Udine, Italy} \\
\r {37} {\eightit University of Tsukuba, Tsukuba, Ibaraki 305, Japan} \\
\r {38} {\eightit Tufts University, Medford, Massachusetts 02155} \\
\r {39} {\eightit Waseda University, Tokyo 169, Japan} \\
\r {40} {\eightit University of Wisconsin, Madison, Wisconsin 53706} \\
\r {41} {\eightit Yale University, New Haven, Connecticut 06520} \\
\end{center}

\pagebreak

\begin{abstract}
  We present a measurement of the
  mass difference $\dmd$ for
  the $\bd$ meson and the statistical power
  of the $b$~flavor tagging methods used.
  The measurement uses 90 ${\rm pb}^{-1}$
  of data from $p\bar{p}$ 
  collisions at $\sqrt{s} = 1.8$ TeV collected with the
  CDF detector.
  An inclusive lepton trigger is used to collect a large sample
  of $B$~hadron semileptonic decays.
  The mass difference $\dmd$
  is determined from the proper time dependence of the fraction of
  $B$ hadrons that undergo flavor oscillations.
  The flavor at decay is inferred from the charge of the
  lepton from semileptonic $B$ decay.
   The initial flavor is inferred by determining the flavor of the
   other $B$~hadron produced in the collision, either from its semileptonic
   decay (soft-lepton tag) or from its jet charge.
  The measurement yields 
  $\Delta m_d = 0.500 \pm 0.052  \pm 0.043$~$\hbar{\rm ps}^{-1}$,
  where the first uncertainty is statistical and the second uncertainty is systematic.
  The statistical powers ($\epsilon D^2$) of the
  soft-lepton and jet-charge flavor taggers are
  $(0.91 \pm 0.10 \pm 0.11)\%$ and
  $(0.78 \pm 0.12 \pm 0.08)\%$, respectively.
\end{abstract}


\pagebreak

\section{Introduction}
In the Standard Model of electroweak interactions~\cite{SM},
the quark mass eigenstates are related to the weak eigenstates
via the unitary $3\times3$ Cabibbo-Kobayashi-Maskawa (CKM) Matrix
$V_{\rm CKM}$~\cite{ckm}.
The nine elements in this matrix,
$V_{ij}$, where $i={\rm u,c,t}$ and $j={\rm d,s,b}$,
are completely determined from three angles and a phase.
A nonzero phase gives \cp~violation
in the weak interaction.
Measurements of decays of hadrons containing $b$~quarks ($B$~hadrons)
are of great interest because they determine the magnitudes
of five of the nine elements of $V_{\rm CKM}$ as well as the phase.

\subsection{The Unitarity Triangle}

The unitarity of $V_{\rm CKM}$ leads to nine unitarity relationships,
one of which is of particular interest:
\begin{equation}
\label{eqn:UT}
V_{\rm ud}V_{\rm ub}^* + V_{\rm cd}V_{\rm cb}^* + V_{\rm td}V_{\rm tb}^* = 0.
\end{equation}
In the complex-plane, this sum of three complex numbers
is a triangle, commonly referred to as {\it the} Unitarity Triangle. 
Measurements of the weak decays of $B$~hadrons and the already known CKM matrix elements
determine the magnitudes of the three sides of the Unitarity Triangle, and
$CP$~asymmetries in $B$~meson decays determine the three angles. 
The primary goal of $B$~physics in the next decade is to measure precisely
both the sides and angles of this triangle and test consistency
within the Standard Model.

We can use several approximations to express eq.~\ref{eqn:UT}
in a more convenient form.
The elements $V_{\rm ud}\simeq 1$ and
$V_{\rm cd}\simeq -\lambda = -\sin\theta_{\rm C}$, where $\theta_{\rm C}$
is the Cabibbo angle, are well measured.
Although the  elements $V_{tb}$ and $V_{ts}$ are not well
measured, the theoretical expectations are that
$V_{tb} \simeq 1$ and $V_{ts} \simeq -V_{cb}^*$.  With
these assumptions, eq.~\ref{eqn:UT} becomes
\begin{equation}
\label{eqn:UTREV}
\frac{V_{\rm ub}^*}{\lambda V_{\rm cb}^*}
-1 - \frac{V_{\rm td}}{\lambda V_{\rm ts}} = 0.
\end{equation}
Measurement of $\dmd$, the subject of this paper, directly impacts the
determination of $V_{\rm td}/V_{\rm ts}$.

\subsection{Determining $\vtd$ and $\vts$ From
Neutral $B$~Meson Flavor Oscillations}

Second-order weak processes transform a neutral $B$~meson into its
antiparticle: $\bd\rightarrow\antibd$, giving a probability for
a $\bd$ to decay as a $\antibd$ that oscillates with time.
The frequency of these oscillations is the mass difference $\dmd$ between
the $B$~mass eigenstates, which are linear combinations of the
flavor eigenstates $\bd$ and $\antibd$.
The mass difference is proportional to $|\vtb^* \vtd|^2$,
so in principle, a measurement of $\dmd$ determines
this product of CKM~matrix elements.
In practice, however, large theoretical
uncertainties limit the precision of $\vtd$.
The same problem exists in determining $\vts$ from $\dms$, the frequency
of $\bs\rightarrow\antibs$ oscillations.
These theoretical uncertainties are reduced in determining the ratio
$|\vtd/\vts|^2$ from $\dmd/\dms$.
Unfortunately, at this time, attempts to measure $\dms$
have only led to lower limits.
The determination of $\dms$ is a key future measurement
of $B$~hadrons, since $|\vtd/\vts|^2$ determines the magnitude of one of the
sides of the Unitarity Triangle, as expressed in eq.~\ref{eqn:UTREV}.  

Measurements of $\dmd$ and $\dms$ require determining the
initial flavor of the $B$~meson, that is, whether the $B$~meson contained
a $b$~quark or a $\bar{b}$~antiquark.
Flavor determination is also crucial in the
measurement of $CP$~violation in the decays
of neutral $B$~mesons to $CP$~eigenstates.

This paper describes a measurement of $\dmd$ using data collected by the CDF
experiment from $\ppbar$ collisions with a center-of-mass energy of 1.8~TeV
produced by the Fermilab Tevatron.
In addition, the measurement of $\dmd$ is used to demonstrate the 
performance of two methods
(described below) of identifying the flavor of B~hadrons in the environment
of $\ppbar$ collisions. 
These methods of flavor identification will be important in the measurement of
\cp~violation in the decays of neutral $B$~mesons and in the study
of $\dms$~\cite{TDR}.

\subsection{Previous Measurements of $\dmd$ from CDF}

The CDF collaboration has exploited the large $b$~quark cross-section at
the Fermilab Tevatron to make several precision measurements of the
properties of $B$~hadrons, including lifetimes and masses~\cite{ref:cdf_b_refs}.
Although the $b$~quark production cross-section at the Tevatron is large,
the $p\bar{p}$ inelastic cross-section is three orders of magnitude larger,
so specialized triggers are required to collect large samples of B~hadrons.
To date, the triggers that have been utilized are based on leptons
(electrons and muons) or dileptons.
Some analyses use the semileptonic decays of $B$~hadrons,
$B\rightarrow\ell\nu X$;
some use the semileptonic decays of charmed particles from $B$~hadron decay
({\it e.g.}~$B\rightarrow D X$, followed by $D\rightarrow\ell\nu Y$);
and some use leptonic $J/\psi$ decays
($B\rightarrow J/\psi X$, followed by $J/\psi\rightarrow\mu^+\mu^-$).

Previous measurements of $\dmd$ from CDF were based on data samples
collected with a dimuon trigger~\cite{ribon} and single-lepton triggers
($\ell = e,\mu$)~\cite{petar}.
The single-lepton triggers were used to partially reconstruct approximately
6,000 $\bd$~mesons via their semileptonic decays
$\bd\rightarrow\ell^+ \nu D^{(*)-} X$ (in this paper, reference to a particular
decay sequence implies the charge-conjugate sequence as well). 
The analysis reported in this paper uses the same data sample collected with this
trigger, but increases the number of $B$ mesons by over an order of magnitude by
inclusively reconstructing $B$~hadrons that decay semileptonically.
The inclusive reconstruction is made possible by the relatively long lifetime
of $B$~hadrons: the decay point of the $B$~hadron is typically separated
from the production point (the primary vertex) by a couple of millimeters.
The inclusive reconstruction is based on identifying this decay point
by associating the trigger lepton with other charged decay products
to reconstruct a secondary vertex.

\subsection{Method of Measuring $\dmd$}

The oscillation frequency $\dmd$ can be found from either a time
independent measurement (that is, from the total number of $\bd$'s
that decay as $\antibd$'s) or from a time dependent measurement
(that is, from the rates that 
a state that is pure $B^0$ at $t=0$ decays as either a $\bar{B^0}$
or $B^0$ as a function of proper decay time $t$).
The latter technique has better sensitivity and allows a
simultaneous study of the tagging methods since the amplitude of
the oscillation depends on the effectiveness of the tagging method.

The expected rate is
\begin{equation}
\label{eqn:pnomix}
{\cal P}(\bd\rightarrow\bd) =
\frac{1}{2\tau_B} e^{-t/\tau_B}( 1 + \cos(\dmd t));
\end{equation}
\begin{equation}
\label{eqn:pmixed}
{\cal P}(\bd\rightarrow\antibd) =
\frac{1}{2\tau_B} e^{-t/\tau_B}( 1 - \cos(\dmd t)),
\end{equation}
where $\tau_B$ is the mean lifetime of the two mass eigenstates of the $\bd$
(the difference in lifetime of these two eigenstates is very small~\cite{PDG98} and has been
neglected), and $t$ is the proper decay time of the $\bd$ in its restframe.
To measure this rate, we need to make three measurements:
(1) the proper decay time, (2) the $B$ flavor at decay, and
(3) the produced $B$~flavor.
We determine the proper time by measuring
the distance from the production
point to the decay point in the laboratory frame
combined with an estimate of the $B^0$ momentum.
The flavor at decay is determined from the charge of the lepton,
assuming it comes from semileptonic $B$ decay. 
In our  measurement~\cite{petar} of $\dmd$ using
$\bd\rightarrow\ell^+ \nu D^{(*)-} X$ decays, the flavor at production was
identified using a same-side tagging technique based on the electric
charge of particles produced in association with the $B^0$.
This flavor tag has also been applied~\cite{sin2beta_prl} to a sample of
$B^0_d/\bar{B}^0_d \rightarrow J/\psi K^0_S$
decays to measure the CP~asymmetry.
To identify the flavor at production in this analysis, we rely on the fact
that the dominant production mechanisms of $b$~quarks in $\ppbar$ collisions
produce $\bbbar$~pairs. 
The flavors of the $B$~hadrons are assumed to be opposite at the 
time of production.
In this paper, we identify the flavor of the other $B$~hadron using two
techniques: the soft-lepton flavor tag (SLT) and
the jet-charge flavor tag (JCT).

Several precise measurements of $\dmd$ have been
published~\cite{MixResults} by experiments operating on the
$\Upsilon(4S)$ and $\z0$~resonances.
The measurement of $\dmd$ presented here
is competitive in precision and in addition quantifies
the performance of the flavor tags, which are crucial for future measurements
of $\cp$~violation in the decays of $B$~mesons 
and the measurement of $\Delta m_s$ at a hadron collider.
The jet-charge flavor tag is a powerful technique in studies of neutral
$B$~meson flavor oscillations by experiments operating on the
$\z0$~resonance~\cite{JetCharge}.
This analysis is the first application of the jet-charge flavor tag in
the environment of a hadron collider.

\section{The CDF Detector}

The data sample used in this analysis was collected from
90~${\rm pb}^{-1}$ of $\ppbar$
collisions recorded with the CDF detector at the Fermilab Tevatron.
The CDF detector is described in detail elsewhere~\cite{detector,TopPRD}.
We summarize here the features of the detector that are
important for this analysis. The CDF coordinate system
has the $z$~axis pointing along the proton momentum, with the $x$~axis
located in the horizontal plane of the Tevatron storage ring, pointing 
radially outward, so that the $y$~axis points up.
The coordinates $r$, $\phi$, and $\theta$ are the standard cylindrical
coordinates.

The CDF spectrometer consists of three separate detectors for tracking charged
particles: the silicon vertex detector (SVX), the vertex detector (VTX),
and the central tracking chamber (CTC), which are immersed in a magnetic
field of 1.4~Tesla pointed along the $+z$~axis.
The SVX~\cite{SVX}
 consists of four concentric cylinders of single-sided silicon strip
detectors positioned at radii between 3~cm and 8~cm 
from the beam line. The strips are oriented parallel to the beam axis
and have a pitch of 60~$\mu$m in the inner three layers and 55~$\mu$m  
on the outermost layer. The SVX is surrounded by the VTX, which is used
to determine the $z$ coordinate of the $\ppbar$ interaction
(the primary vertex).
Surrounding the
SVX and VTX is the CTC. The CTC is a drift chamber that is 3.2~m long with
84 layers of sense wires located between a radius of 31~cm and 133~cm.
The sense wires are organized into five axial superlayers and four stereo
superlayers with a stereo angle of $3^\circ$. The momentum resolution of
the spectrometer is
 $\delta \pt/\pt=[(0.0009{\rm GeV}/c\cdot\pt)^2 + (0.0066)^2]^{\frac{1}{2}}$,
where $\pt$ is the component of momentum transverse to the $z$~axis
($\pt = p\cdot\sin\theta$). 
Charged particle trajectories reconstructed
in the CTC that are matched to strip-clusters in the SVX have an impact
parameter resolution of
$\delta \dzero = \left[13 + (40 \ {\rm GeV}/c)/\pt\right] \ \mu$m,
where the impact parameter
$\dzero$ is the distance of closest approach of the trajectory to the
beam axis in the plane perpendicular to the beam axis.
The outer 54 layers of the CTC are instrumented to record the
ionization ($dE/dx$) of charged tracks.

Surrounding the CTC are the central electromagnetic calorimeter (CEM)
and the central hadronic calorimeter (CHA).  
The CEM has strip chambers (CES) positioned at shower maximum,
and a preshower detector (CPR) located at a depth of one radiation length.
Beyond the central calorimeters lie two sets of muon detectors.
To reach these two detectors, particles produced at the primary vertex
with a polar angle of $90^\circ$ must traverse material totaling
5.4 and 8.4 pion interaction lengths, respectively.
The trigger system consists of three levels: the first two levels are
implemented in hardware. 
The third level consists of software reconstruction algorithms
that reconstruct the data, including three-dimensional track
reconstruction in the CTC using a fast algorithm
that is efficient only for $\pt>1.4$~\gevc.


\section{Data Sample Selection}

The sample selection begins with data from the inclusive $e$ and $\mu$
triggers.
At Level~2, both of these triggers require a track
with $\pt>7.5$~\gevc~found by the central fast tracker (CFT)~\cite{cft},
a hardware track processor that uses fast timing information from the
CTC as input.
The resolution of the CFT is
$\delta \pt/\pt=0.035 ({\rm GeV}/c)^{-1}\cdot\pt$.
In the case of the electron trigger, the CFT track must be matched to a
cluster in the electromagnetic calorimeter, with transverse energy
$\et>8.0$~GeV, where $\et=E\cdot\sin\theta$, and  $E$ is the energy
of the calorimeter cluster.
In the case of the muon trigger, the CFT track must be
matched to a reconstructed track-segment in both sets of muon detectors.
In the third level of the trigger, more stringent electron and muon
selection criteria, which are similar to the selection criteria
described in Section~\ref{sec:lepton_id}, are applied.
The inclusive electron data set contains approximately
5.6~million events and the inclusive muon data set contains
approximately 2.0~million events.
These data are dominated by leptons from the decay of heavy flavors
($b\rightarrow\ell$ and $c\rightarrow\ell$) and hadrons that mimic
the lepton signal.

\subsection{Electron and Muon Identification}
\label{sec:lepton_id}

Electron candidates are identified using information from both the
calorimeters and the tracking detectors.
The electron calorimeter cluster in the CEM must have $E_T>6$~GeV.
The longitudinal shower profile of this cluster is required to be
consistent with an electron shower with a leakage energy from the CEM
into the CHA of less than 4\%. 
The lateral shower profile of the CEM
cluster has to be consistent with the profile determined from test beam electrons.
A track with $p_t> 6$~GeV/$c$ must match the electron
calorimeter cluster.
This match is based on a comparison of the track position
with the calorimeter cluster position determined in the CES:
the difference between the extrapolated position of the track and the position of
the cluster centroid must satisfy $r|\Delta \varphi| < 1.5$~cm and
$|\Delta z\sin \theta| < 3$~cm.

To identify muons, we require a match between the extrapolated
CTC track and the track segment in the muon chamber in both
the $r$-$\varphi$ and $r$-$z$ view.
The uncertainty in this match is taken into account and is dominated
by multiple scattering in the detector material.
The transverse muon momentum must satisfy $p_T > 6$~GeV/$c$. 

Finally, to ensure optimal resolution of the $B$~hadron decay point,
the electron and muon candidate tracks have to be reconstructed in the
SVX detector. 

\subsection{Jet Reconstruction}

Further analysis of the data sample is based on the charged particle jets
in the event.
Charged particles (instead of the more commonly used calorimeter clusters)
are used to form jets in order to keep the electron and muon samples as
similar as possible. 
These jets are found using a cone clustering algorithm.
Tracks with $\pt>1.0$~\gevc\ are used as jet seeds.
If two seeds are within $\Delta R < 0.7$~\cite{deltaR},
the momenta of the seeds are added together to form
a new seed. 
After all possible seed merging, lower momentum tracks ($0.4<\pt<1.0$~\gevc) 
that are within $\Delta R<0.7$ of a seed are added in to form the final
jets.
The trigger lepton is always associated to a jet, and below we refer to this
jet as the trigger-lepton jet.
A jet can consist of a single track with $\pt>1$~\gevc.


\subsection{Secondary Vertex Reconstruction}

In order to reconstruct the time of decay in the $B$ rest frame
(the proper time), we must measure the point of decay 
with respect to the primary interaction in the lab
and estimate the momentum of the $B$.
Since the SVX provides only coordinates in the plane transverse to the
beam axis, the measurement of the separation between the
point of decay and the primary vertex is done only in the
$x$-$y$ plane.  
We refer to this separation as the decay length $L_{xy}$.  
Only the component of the $B$ momentum transverse to the
beam axis ($\ptb$) is needed to calculate the proper
time at decay, since the decay length is measured in
the $x$-$y$ plane.

The positions of the $p\bar{p}$ interactions or ``primary vertices''
are distributed along the beam direction according to a Gaussian
with a width of $\sim 30$~cm.
In the plane transverse to the beam axis, these interactions follow a
distribution that is a Gaussian  with a width 
of $\sim 25~\mu$m in both the $x$ and $y$ dimensions.
To reconstruct the primary event vertex, we first identify its
$z$-position using the tracks reconstructed in the VTX detector. 
When projected back to the beam axis, these tracks determine the longitudinal
location of the primary interaction with a precision of about
0.2~cm along the beam direction.
If there is more than one reconstructed primary vertex in an event,
the trigger lepton is associated with the primary vertex closest in $z$
to the intercept of the trigger lepton with the beam line.

The transverse position of the primary vertex is determined for each event by a
weighted fit of all tracks with a $z$ coordinate within 5~cm of the $z$-vertex
position of the primary vertex associated with the trigger lepton.
The tracks used in this fit are required to have been reconstructed 
in the SVX detector. 
First all tracks are forced to originate from a common vertex.
The position of this vertex is constrained by the transverse beam envelope
described above.
Tracks that have large impact parameters with respect to this vertex
are removed, and the fit is repeated.
This procedure is iterated until all tracks are within the
required impact parameter requirement.
At least five tracks must be used in the determination of the transverse
position of the primary vertex or 
we use the nominal beam-line position.
The primary vertex coordinates transverse to the beam direction 
have an uncertainty in the range
10-35~$\mu$m, depending on the number of tracks and the event topology.

The reconstruction of the $B$ decay point (referred to below as the
secondary vertex) in the trigger-lepton jet
is based on the technique developed to identify jets formed by
$b$~quarks coming from $t$~quark decay~\cite{TopPRD}.
Some modifications to this technique were necessary to maintain
good efficiency for reconstructing the $B$~hadron decay point
in our data sample, since the $B$~hadrons in this sample have substantially
lower $\pt$ than the $B$~hadrons from top~quark decay.
The search for a secondary vertex in the trigger-lepton jet is
a two stage process.
In both stages, tracks in the jet are selected for reconstruction
of a secondary vertex based on the significance of their impact parameter
with respect to the primary vertex,
$d_0/\sigma_{d_0}$, where $\sigma_{d_0}$ is the estimate of the uncertainty
on $d_0$. The uncertainty $\sigma_{d_0}$ includes contributions from both the
primary vertex and the track parameters. The first stage requires at least
three candidate tracks for the reconstruction of the secondary vertex.
The trigger lepton is always included as a candidate, whether or not
it satisfies the $d_0/\sigma_{d_0}$ requirement.
Tracks consistent with coming from the decay $K^0_S\rightarrow \pi^+\pi^-$
or $\Lambda^0\rightarrow p\pi^-$ are not used as candidate tracks.
Two candidate tracks are constrained to pass through the same space point
to form a seed vertex. If at least one additional candidate track
is consistent with intersecting this seed vertex, then the seed vertex is used as
the secondary vertex.
If the first stage is not successful in finding a secondary vertex,
the second stage is attempted. More stringent track requirements
(on $d_0/\sigma_{d_0}$ and $\pt$, for example) are imposed on the
candidate tracks. All candidate tracks satisfying these stricter criteria
are constrained to pass through
the same space point to form a seed vertex. This vertex has an associated
$\chi^2$. Candidate tracks that contribute too much to the $\chi^2$ 
are removed, and a new seed vertex is formed. This
procedure is iterated until a seed vertex remains that has at least two
associated tracks and an acceptable value of $\chi^2$.
The trigger lepton is one of the tracks used to determine the
trigger-lepton jet secondary vertex in 96\% of the events.

\begin{figure}
\postscript{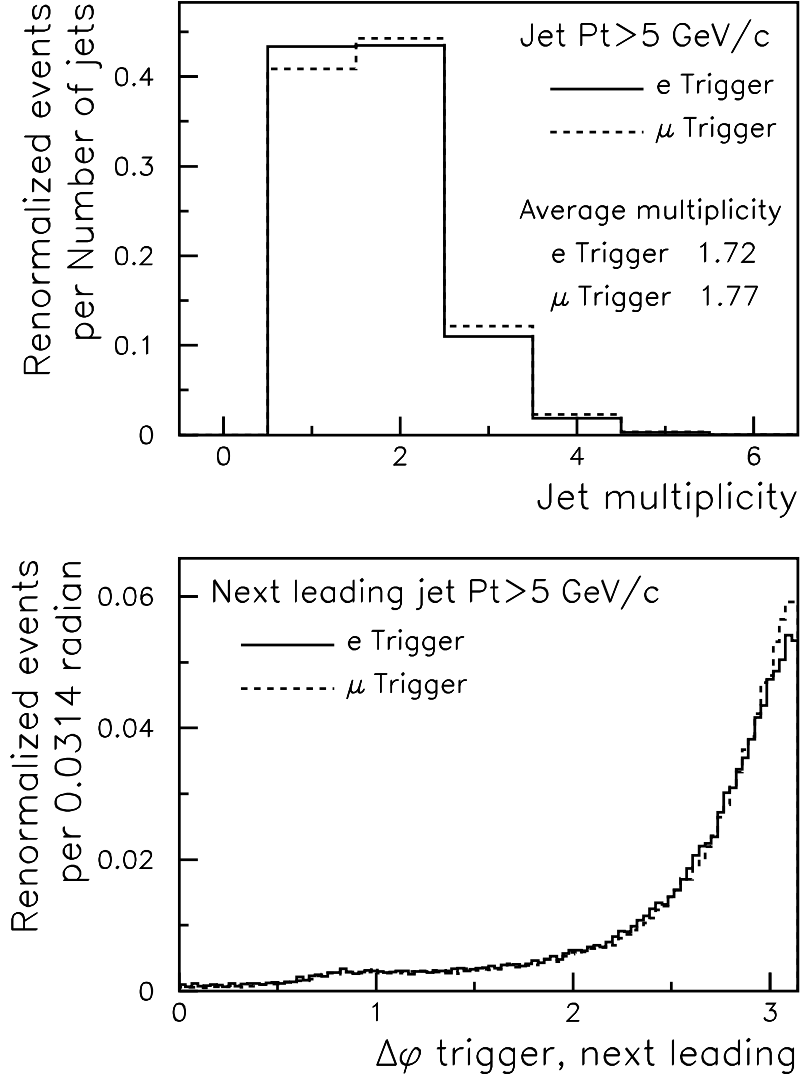}{1.0}
\caption{ 
Jets distributions from the inclusive lepton data samples
after the trigger lepton has been associated with a secondary vertex.
The upper figure shows the
number of jets with transverse momentum $p_t>5$~GeV/$c$ per event.
Approximately 60\% of the events have more than one jet. The lower
figure shows the separation in azimuth of the trigger-lepton jet
and the jet with the largest $p_t$ in these events.
}
\label{fig:jets}
\end{figure}

The decay length of the secondary vertex $L_{xy}$ is the projection of
the two-dimensional vector pointing from the primary vertex to the secondary
vertex on the jet axis (defined by the sum of all the momenta of the tracks
included in the jet);
if the cosine of the angle between these two vectors is positive (negative),
then $L_{xy}$ is positive (negative). 
Secondary vertices from the decay of
$B$~hadrons are expected to have positive $L_{xy}$, while negative
$L_{xy}$ vertices usually result from random combinations of
mismeasured tracks.
To reduce the background from these false vertices,
we require $|L_{xy}/\sigma_{L_{xy}}| > 2.0$,
where $\sigma_{L_{xy}}$ is the estimated uncertainty on $L_{xy}$.
We require a secondary vertex to be associated with
the trigger-lepton jet. This requirement leaves us with
243,800 events: 114,665 from the electron data sample
and 129,135 from the muon data sample.
The fraction of events with $L_{xy}<0$ is 4.5\% in the electron
data sample and 5.7\% in the muon-trigger sample.
For reasons we discuss later, only events with $L_{xy}>0$
are used in the determination of $\Delta m_d$ and the study
of the performance of the flavor tags.
The distribution of the number of jets with total transverse
momentum $\pt>5$~GeV/$c$ is shown in the upper plot of
Figure~\ref{fig:jets}. Approximately 60\% of the events contain
a second jet in addition to the trigger-lepton jet.
The lower plot in Figure~\ref{fig:jets} shows the difference in azimuth 
between the trigger-lepton jet
and the jet with the highest $\pt$ in these events. 
A large fraction of these jets are back-to-back with the trigger-lepton
jet as expected from the lowest-order processes that produce $b\bar{b}$
pairs.
We search for secondary vertices in the other jets in the events
as well. If an additional secondary vertex is found in one of these
other jets, we classify this event as a ``double-vertex'' event.
If only the single secondary vertex associated with the trigger-lepton
jet is found, this event is classified as a ``single-vertex'' event.
The distinction between single-vertex and  double-vertex events is important
in applying the jet-charge flavor tag as described below.
\subsection{Determination of the $B$ Hadron Flavor} 

The next step in the analysis is to identify the flavor at production of the $B$~hadron
that produced the trigger lepton.
We accomplish this by identifying the flavor of
the other $B$~hadron produced in the collision. 
We refer to this other $B$~hadron as the ``opposite~$B$'' in the text below.
We first search for an additional lepton coming from the semileptonic decay
of this opposite~$B$. 
Because the $\pt$ of this lepton is 
not biased by the trigger, it is typically
much smaller than the $\pt$ of the trigger lepton, so 
we call this method of flavor identification the ``soft-lepton tag'' or SLT.
The soft lepton can be either an electron or a muon. 
The lepton selection
criteria are similar to the selection criteria described in 
Section~\ref{sec:lepton_id} and Reference~\cite{TopPRD},
with additional selection criteria that use $dE/dx$,
and pulse height in the CPR and CES.
The soft lepton must have a track with $\pt>2$~GeV/$c$, and the invariant mass
of the soft lepton and the trigger lepton must be greater than 5~GeV/$c^2$.
This requirement removes soft leptons coming from sequential semileptonic
decays of charm particles produced in the decay of the $B$~hadron producing
the trigger lepton. The lepton identification criteria restrict the
acceptance of the soft leptons to a pseudorapidity of $|\eta|<1.0$,
where \mbox{$\eta = - \ln( \tan( \frac{\theta}{2}))$}.
Approximately 5.2\% of the 243,800 events contain a soft lepton candidate.

If a soft lepton is not found, we try to 
identify the jet produced by the opposite~$B$.
We calculate a quantity called the jet charge $Q_{\rm jet}$ of this jet:
\begin{equation}
  Q_{\rm jet} = \frac{ \sum_i q_i \cdot (\vec{p}_i \cdot\hat{a}) }
                     { \sum_i \vec{p}_i \cdot\hat{a}  },
\end{equation}
where $q_i$ and $\vec{p}_i$ are the charge and momentum
of the $i^{\rm th}$ track in the jet and $\hat{a}$ is
a unit-vector defining the jet axis.
For $b$-quark jets, the sign of the jet charge is on average the same as the sign
of the $b$-quark that produced the jet, so the sign of the jet charge
may be used to identify the flavor at production of the $B$~hadron producing
the trigger lepton.
If a second jet in the event other than the trigger-lepton jet has
a secondary vertex, then we use this jet to calculate the jet charge.
Double-vertex events with a jet-charge flavor tag are referred to
as JCDV events. 
If only the trigger-lepton jet contains
a secondary vertex, we search for a jet
with $\Delta \phi > \pi/2$ with respect
to the trigger lepton and $\pt>5$~\gevc.
If there is more than one jet satisfying the above
criteria, we choose the jet with the highest
$\pt$.
Single-vertex events with a jet-charge flavor tag
are referred to as JCSV events.
Approximately 7.5\% of the 243,800 events above are JCDV events
and approximately 42\% are JCSV events.


\section{Data Sample Composition}

The events in our selected data sample come from three sources:
$b\bar{b}$~production, $c\bar{c}$~production, and  
light quark or gluon production. 
In each event, the trigger lepton may be a true lepton or
it may be a hadron that mimics the experimental signature of a lepton
(a fake lepton).
The secondary vertex in the trigger-lepton jet may be a true vertex 
due to the decay of heavy flavor ($b$ or $c$)
or a random combination of erroneously reconstructed tracks
that appear to form a vertex that is displaced
from the primary interaction (a fake vertex).
Light quark or gluon jets produce false vertices with $L_{xy}>0$
with equal probability as false vertices with $L_{xy}<0$.
The small fraction ($<$6\%) of events with $L_{xy}<0$ indicates that
this background is small.
In the analysis, we assume that all events come from
heavy flavor ($b\bar{b}$ and $c\bar{c}$) production.
The probability of a light quark or gluon
event producing a fake vertex and a fake lepton
is negligible,
although in the evaluation of the systematic uncertainties,
we take into account the possible effects of a small
amount of non-heavy flavor background.
Below, we describe how we determine the fraction of our samples that
are due to $b\bar{b}$ production, $c\bar{c}$ production, and fake leptons.
\subsection{Simulation of Heavy Flavor Production and Decay}

To understand the composition of our data, we use Monte Carlo samples of
$b\bar{b}$ and $c\bar{c}$ production.
Version 5.6 of the {\tt PYTHIA}~\cite{Pythia} Monte Carlo generator
was used to generate high-statistics $b\bar{b}$ and $c\bar{c}$ samples.
The $b\bar{b}$ and $c\bar{c}$ pairs are generated through processes
of order up to $\alpha_s^2$ such as $g g \rightarrow \mathbf{q} \bar{\mathbf{q}}$ and
$q \bar{q} \rightarrow \mathbf{q} \bar{\mathbf{q}}$,
where $\mathbf{q} =$~$b$~or~$c$.
Processes of order $\alpha_s^3$, such as gluon splitting,
where $g g \rightarrow g g$ is followed by
$g \rightarrow \mathbf{q} \bar{\mathbf{q}}$, are not included,
but initial and final state gluon radiation is included.
The $b$ and $c$~quarks are hadronized using the Peterson~\cite{Peterson}
fragmentation function with the parameters
$\epsilon_b=0.006$ and $\epsilon_c=0.06$.
The bottom and charm~hadrons were decayed using version 9.1 of the
CLEO Monte Carlo {\tt QQ}~\cite{QQ}.
Events with a lepton with $\pt>6$ GeV/$c$ were accepted based on 
an efficiency parameterization~\cite{ref:cft_eff_param} of the CFT trigger
that depends on the lepton $\pt$.
The accepted events were passed through a simulation of the CDF detector
that is based on parameterizations and simple models
of the detector response that are functions of the particle kinematics.
After the simulation of the CDF detector, the Monte Carlo events were
treated as if they were real data.

\subsection{Sources of Trigger-Electrons}

The trigger-electrons in the sample can come from three sources:
heavy flavor decay
($b\rightarrow e$, $b\rightarrow c\rightarrow e $, and $c\rightarrow e$), 
photon conversion
($\gamma \rightarrow e^+e^-$ or $\pi^0\rightarrow \gamma e^+e^-$),
or hadrons that fake the electron signature in the detector.
The contribution from heavy flavor decay is discussed in
section~\ref{sec:heavy_flavor}.
We attempt to identify and reject photon conversions by searching for the
partner of the trigger-electron. We search for an oppositely charged track
that forms a good, zero opening angle vertex with the trigger-electron.
The $dE/dx$ of this track, as measured in the CTC,
must be consistent with the electron hypothesis.
We removed 2\% of the electron trigger sample that was identified as
photon conversions.
We estimate that about 1\% of the remaining events
contain a trigger-electron from a photon conversion
that was not identified.
To determine the fraction of events that contain a
hadron that fakes an electron,
we fit the trigger-electron $dE/dx$ spectrum 
for its $e$, $\pi$, $K$, and $p$ content.
We found the non-electron fraction of the sample 
to be (0.6 $\pm$ 0.5)\%, where the uncertainty is statistical
only.
Since this background is small, we neglect it in the remainder of the analysis.

\subsection{Sources of Trigger-Muons}
\label{sec:fakeMu}

The trigger muons in the sample can come from
heavy flavor decay
($b\rightarrow\mu$, $b\rightarrow c\rightarrow\mu$, and $c\rightarrow\mu$),
$\pi$ and $K$ decay, and from
hadrons that penetrate the absorbing material
in front of the muon chambers.
The contribution from heavy flavor decay is discussed in
section~\ref{sec:heavy_flavor}.
To study the properties of fake muon events, we used a control sample of events
that only required a high $\pt$ CFT track in the trigger.
The trigger track was treated like a trigger lepton, and
the jet containing the trigger track was required
to contain a secondary vertex.
The $L_{xy}$ distribution of the control sample
was very similar to the heavy flavor Monte Carlo
$L_{xy}$ distributions and the $L_{xy}$ of the signal data samples.
We conclude from this comparison that
due to the secondary vertex requirement
most of the fake muon events in the data
are events from heavy flavor production.
As described in Appendix~\ref{app:fake_muons},
we estimate the fraction of events with a fake
trigger muon whose dilution is zero by comparing the
flavor tagging performance of the $e$ and $\mu$ trigger
data. The estimated fraction of events with fake muons 
that have zero dilution is $(12\pm 6)$\%.

\subsection{Fraction of Data Sample due to Heavy Flavor Production and Decay}
\label{sec:heavy_flavor}

We determine the fraction of events in the data due to
$b\bar{b}$ and $c\bar{c}$ production
using two kinematic quantities:
the trigger-lepton $\ptrel$ and the invariant mass $\mcl$ of the cluster of
secondary vertex tracks.  
The quantity $\ptrel$ is defined as the magnitude of the 
component of the trigger-lepton momentum that is
perpendicular to the axis of the trigger-lepton jet.
The trigger lepton is removed from the jet, and the jet axis
is recalculated to determine $\ptrel$.
To calculate $\mcl$, we assign the pion mass to all of the tracks
used to form the secondary vertex (except the trigger lepton). 
We include the trigger lepton even
if it is not attached to the secondary vertex.
These kinematic quantities are effective in discriminating between $b\bar{b}$
and $c\bar{c}$ events because of the significant
mass difference between the hadrons containing $b$ and $c$~quarks
($\approx 3$~GeV/$c^2$).

\begin{figure}
\postscript{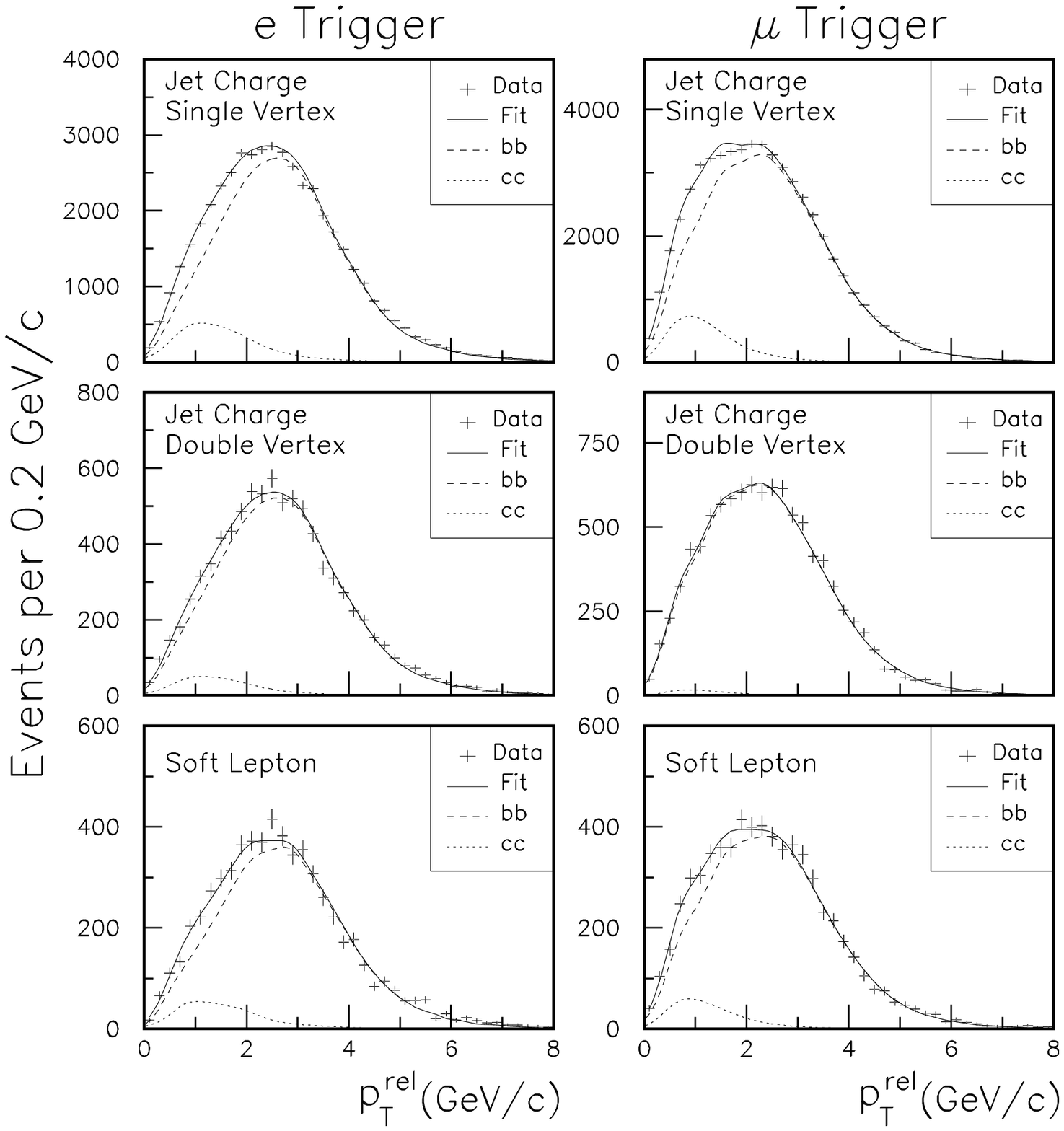}{1.0}
\caption{ 
The sample composition fits using $\ptrel$. The left hand plots are
for the electron data and the right hand plots are for the muon data.
The fit values for the fraction of events from $b\bar{b}$ production
are given in Table~\ref{tab:mclptrel}.
}
\label{fig:samp_comp_ptrel}
\end{figure}

\begin{figure}
\postscript{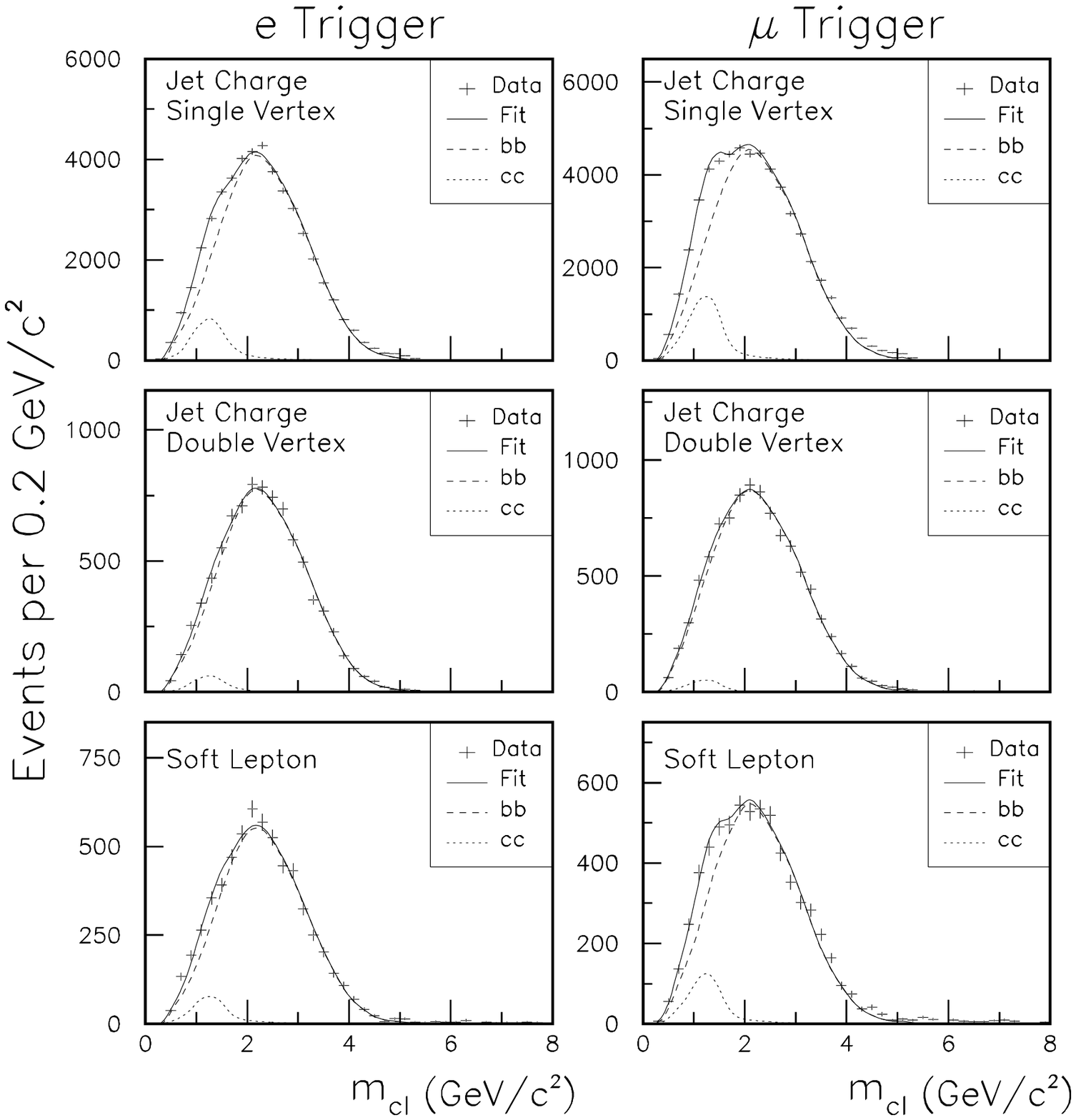}{1.0}
\caption{
The sample composition fits using $m_{\rm cl}$. The left hand plots are
for the electron data and the right hand plots are for the muon data.
The fit values for the fraction of events from $b\bar{b}$ production
are given in Table~\ref{tab:mclptrel}.
}
\label{fig:samp_comp_mcl}
\end{figure}

Template $\ptrel$ and $\mcl$ distributions were obtained from
the $b\bar{b}$ and $c\bar{c}$ Monte Carlo samples.
The $\ptrel$ and $\mcl$ distributions for the data were fit
to the sum of the $b\bar{b}$ and $c\bar{c}$ Monte Carlo templates,
where the normalization for each template was a free parameter.
The $e$ and $\mu$ trigger data were fit separately,
and the data for each trigger were divided according to
flavor tag.
The three categories were: soft-lepton (SLT), jet-charge 
single-vertex (JCSV), and jet-charge double-vertex (JCDV).
The results of the $\mcl$  and $\ptrel$ fits were
averaged to obtain the nominal values for the
fraction of events from  $b\bar{b}$  production 
($F_{b\bar{b}}$).
The fits are shown in
Fig.~\ref{fig:samp_comp_ptrel} and Fig.~\ref{fig:samp_comp_mcl},
and Table~\ref{tab:mclptrel}
gives the nominal values of $F_{b\bar{b}}$ 
for the $e$ and $\mu$ trigger data.
The data are mostly ($>90\%$) from $b\bar{b}$ 
production.

\section{Method of Measuring the Flavor Tag $\epsilon D^2$ and $\Delta m_d$}

As outlined in the introduction, to measure $\Delta m_d$ we compare the
flavor of the $B^0$~meson when it was produced to the flavor of the
$B^0$~meson when it decays as a function of the proper decay time of the
meson. 


\subsection{Reconstruction of the $B$ Proper Decay Time $t$}

To reconstruct the time of decay 
in the $B$ rest frame ($t$), we must combine the two-dimensional
decay length ($L_{xy}$) with the component of the $B$ 
momentum in the $x$-$y$ plane ($\ptb$).  
The proper time is
\begin{equation}
\label{eqn:ctau}
  t = \frac{ L_{xy} \cdot m_B }{ c \cdot \ptb }
\end{equation}
where $m_B$ is the mass of the $B^0$ and $c$ is the
speed of light.
The proper decay length
is the proper time multiplied by the speed of light ($ct$).
We do not observe all of the decay products of the 
$B$:  
the neutrino from the semileptonic decay is not detected,
as well as other neutral decay products and charged decay products
that may not have been associated with the secondary vertex.
This means that $\ptb$ is not known and must be estimated based on  
observed quantities and the $b\bar{b}$ Monte Carlo.

\begin{figure}
\postscript{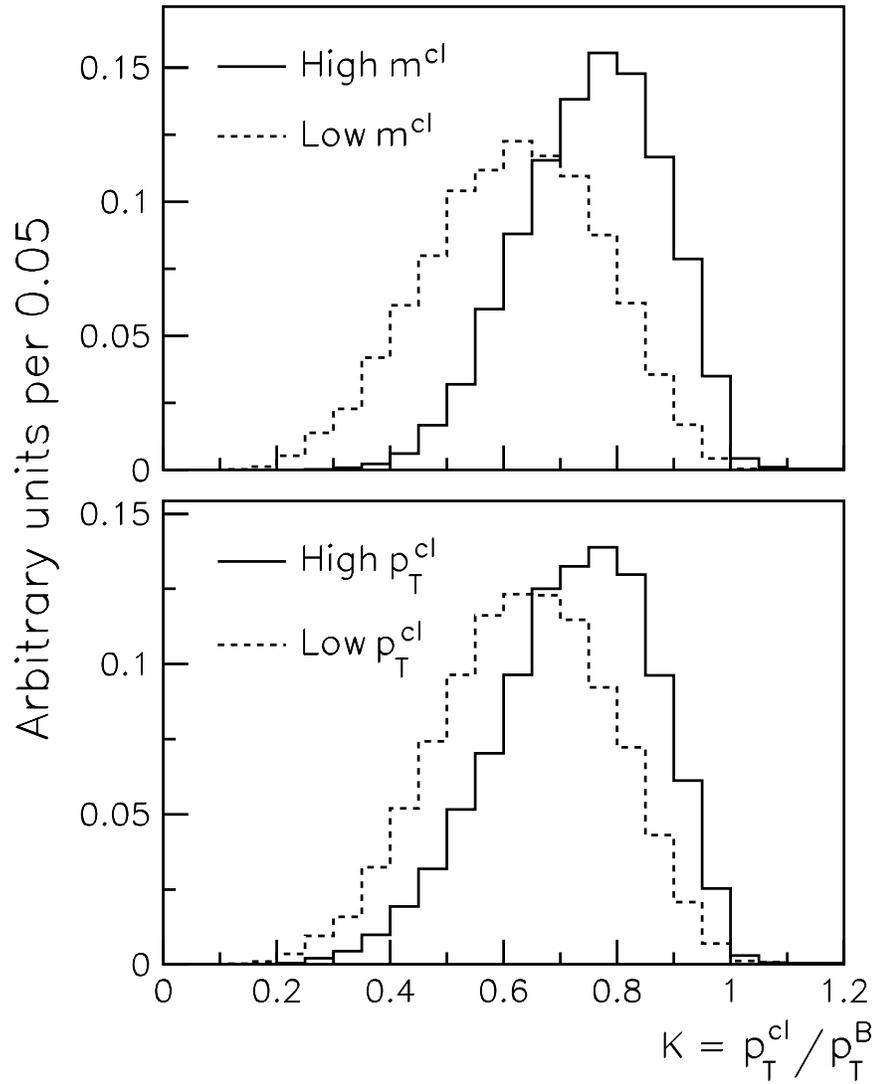}{1.0}
\caption{
   The observed momentum fraction, which
   is defined as $p_T^{\rm cl}/p_T^B$.  The top plot shows 
   two ranges of cluster mass ($m^{\rm cl}$).  The bottom
   plot shows two ranges of cluster transverse momentum
   ($p_T^{\rm cl}$).  The plots were obtained from the $e$-trigger
   $b\bar{b}$ Monte Carlo sample.
}
\label{fig:Kfac}
\end{figure}

The momentum in the transverse plane of the cluster
of secondary vertex tracks $\ptcl$ and $\mcl$ are
the observed quantities used in the estimation of $\ptb$.  
The trigger lepton is included in the calculation of $\ptcl$ and
$\mcl$, even if it is not attached to the secondary vertex, since we
assume that it is a $B$ decay product.
The estimate of the $B$ hadron transverse momentum
$p_T^{B\prime}$ for an event is 
\begin{equation}
   p_T^{B\prime} = \frac{\ptcl}{\langle K \rangle}
\end{equation}
where $\langle K \rangle$ is the mean of 
the distribution of $K=\ptcl/\ptb$ determined
with the $b\bar{b}$ Monte Carlo.
Figure~\ref{fig:Kfac} shows $K$~distributions
for two ranges of $\mcl$ and two ranges of $\ptcl$
in the $e$-trigger $b\bar{b}$ Monte Carlo sample.
For higher $\mcl$ and $\ptcl$ values, the $K$~distribution
has a higher mean and a narrower width: a larger fraction of the
$B$~momentum is observed so the observed $\ptcl$
is a more precise estimate of the $B$~momentum.
To take advantage of the $\mcl$ and $\ptcl$ dependence of the
$K$~distribution, we bin the data in four ranges of $\mcl$ and $\ptcl$ 
for a total of 16 $K$~distributions.
Different sets of $K$~distributions are used for the 
$e$ and $\mu$ trigger data.

\begin{figure}
\postscript{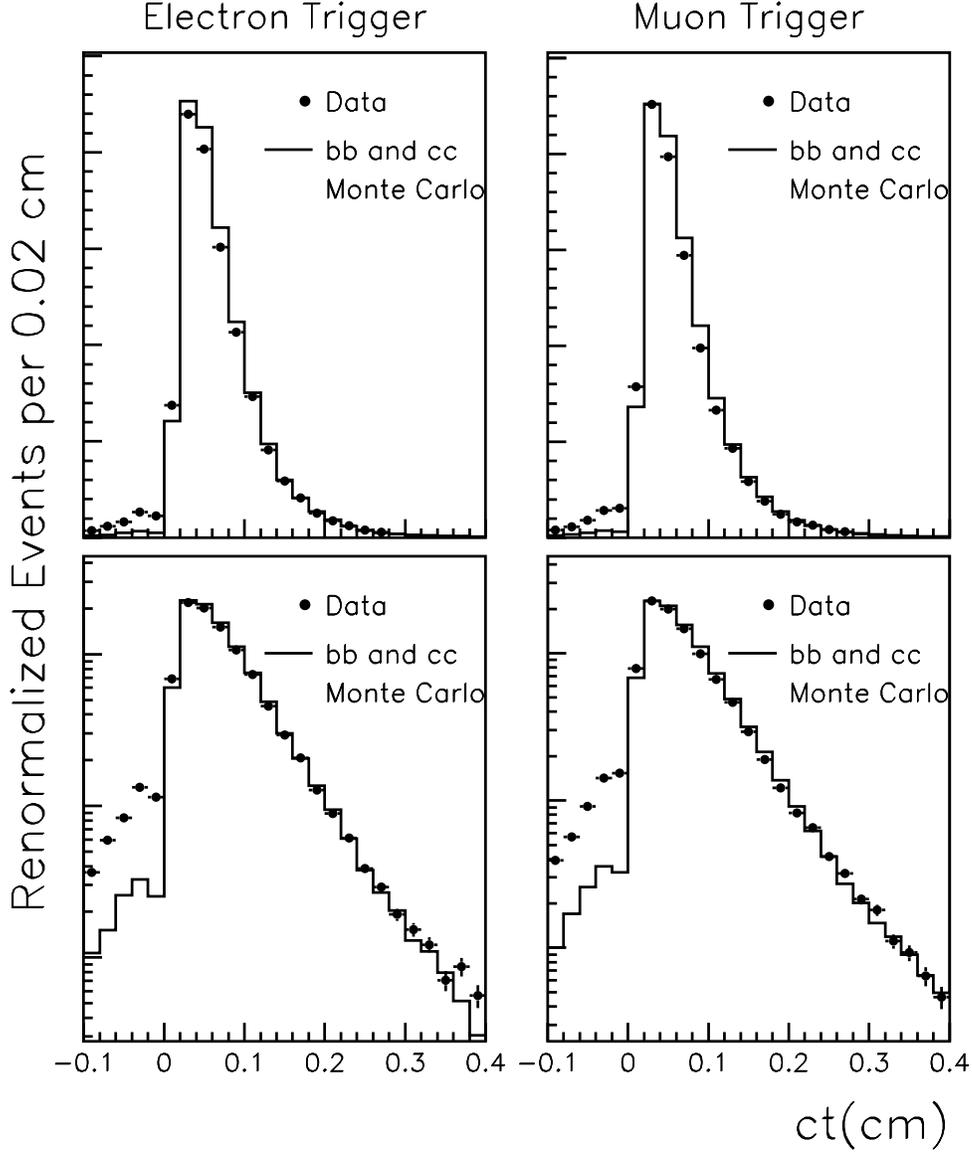}{1.0}
\caption{
   The reconstructed proper decay length $ct$ for the $e$ and
   $\mu$ trigger data.  The points are the data.  The
   solid histogram is a combination of $b\bar{b}$ and $c\bar{c}$
   Monte Carlo with the relative fractions given in Table~\ref{tab:mclptrel}.  
   All distributions are normalized to unit area.  The upper plots show the data
   on a linear scale while the lower plots show it on a logarithmic scale.
}
\label{fig:ctau}
\end{figure}

Figure~\ref{fig:ctau} shows the reconstructed proper
decay length ($ct$) distributions for the $e$ and $\mu$ trigger data.
The plots of the data are compared to the expected 
shape from $b\bar{b}$ and $c\bar{c}$ production,
where the $b\bar{b}$ and $c\bar{c}$ distributions were
combined using $F_{b\bar{b}}$ in Table~\ref{tab:mclptrel}
and setting the fraction of $c\bar{c}$ events 
to $F_{c\bar{c}}=1-F_{b\bar{b}}$.
The lack of events near $ct=0$ is due to the transverse
decay length significance requirement $|L_{xy}/\sigma_{L_{xy}}|>2$.
The exponential fall-off of the data in $ct$ agrees 
with the Monte Carlo prediction. 
There is, however, an excess of events with $ct<0$ in the data.
This excess is due in part to backgrounds and higher-order processes
not included in the Monte Carlo.
These backgrounds include electrons from photon conversions and fake muons.
The higher-order processes include gluon splitting to $\bbbar$ pairs,
which can produce a pair of $B$~hadrons that are close in $\Delta R$.
In these events, the decay products of both $B$~hadrons may be included
in the same jet.
This leads to secondary vertices that include tracks from both $B$~hadrons,
resulting in a specious measurement of $L_{xy}$ that may be positive or
negative.
To verify this, we generated a sample of $\bbbar$ events using the
{\tt ISAJET}~\cite{Isajet} event generator, which includes gluon splitting.
This sample showed an increased fraction of reconstructed vertices
with negative $L_{xy}$.

\subsection{Determination of the $B$ Flavor at Decay}

   The flavor of the $B$ hadron associated with the trigger lepton
at the time of decay is identified by the trigger-lepton charge,
assuming the lepton is from a semileptonic $B$ decay.  
A $B$ hadron that contains an anti-$b$ quark
($\bar{b}$) will give a positively charged lepton in a
semileptonic decay.
The trigger lepton can also originate from the decay of 
a charmed hadron produced in the decay of the $B$~hadron,
{\it e.g.}~$B \rightarrow D X$,$D \rightarrow \ell^- X$,
which produces ``wrong-sign'' trigger leptons,
or $B \rightarrow J/\psi X$, $J/\psi \rightarrow \ell^+\ell^-$,
which produces both wrong-sign and right-sign trigger leptons.
We refer to trigger leptons from these sources as sequential
leptons.
The Monte Carlo $b\bar{b}$ sample is used to determine the
fraction of trigger leptons, $f_{\rm seq}$, from these sequential decays.
We find $f_{\rm seq}$ is 9.4\% in the electron sample and
13.6\% in the muon sample. Approximately 75\% of these
sequential decays, {\it i.e.}, 7.0\% and 10.2\%,
produce wrong-sign leptons with the charge opposite
to the charge from direct semileptonic decay.
We assign a systematic uncertainty of 25\% of its value
to the fraction of sequential leptons,
based on uncertainties on the branching fractions included in 
the CLEO Monte Carlo {\tt QQ} and on measurements of these branching
fractions at the $\Upsilon(4S)$ and the $Z^0$~resonance.

\subsection{Determination of the $B$ Flavor at Production}
To determine the flavor of the trigger-lepton $B$ at the time of production, 
we attempt to identify the flavor of the other $B$ in the event,
and assume that the original flavor of the trigger-lepton $B$
is opposite that of the other $B$.  
As described previously,
we use two methods to obtain the flavor of the other $B$ in
the event: soft-lepton tagging (SLT) and jet-charge tagging (JCT).
The jet-charge tag has two sub-classes:
jet-charge double-vertex (JCDV) and  jet-charge single-vertex (JCSV).
The soft-lepton method is the most effective
({\it i.e.}, has the highest probability of producing a correct tag), but least efficient
method.  
The jet-charge methods are less effective, but more
efficient.  
The presence of a secondary vertex in the jet used for 
the jet charge greatly enhances its effectiveness.

\subsubsection{Quantifying the Statistical Power of the Flavor Tags}

We quantify the statistical power of the flavor tagging
methods with the product 
$\epsilon D^2$, where $\epsilon$ is the
efficiency for applying the flavor tag,
and $D$ is the dilution, which is related
to the probability that the tag is correct
($P_{\rm tag}$) by 
\begin{equation}
\label{eqn:trueD}
  D = 2 \cdot P_{\rm tag} - 1.
\end{equation}
We measure $\epsilon D^2$ in our data.
To illustrate the statistical significance of the product $\epsilon D^2$,
we discuss an asymmetry measurement with two types of events, $a$ and $b$,
where the flavor tagging method identifies whether the
event is of type $a$ or type $b$.
Type $a$ and type $b$ could be ``mixed'' and ``unmixed'' decays
of a neutral $B$~meson, for example.
The measured asymmetry~$A_{\rm meas}$ is
\begin{equation}
  A_{\rm meas} = \frac{N_a - N_b}{N_a + N_b},
\end{equation}
where $N_a$ and $N_b$ are the number of events that are
tagged as type $a$ and type $b$, respectively.
The true asymmetry~$A$ is
\begin{equation}
  A = \frac{N_a^0 - N_b^0}{N_a^0 + N_b^0},
\end{equation}
where $N_a^0$ and $N_b^0$ are the true number of events of type
$a$ and type $b$, respectively, in the sample.
The efficiency is
\begin{equation}
\epsilon=\frac{N_a + N_b}{N_a^0 + N_b^0}.
\end{equation}
The true asymmetry is related to the measured asymmetry by
\begin{equation}
 A = \frac{1}{D} A_{\rm meas};
\end{equation}
and the statistical uncertainty on the true asymmetry is
\begin{equation}
  \sigma_A = \sqrt{ \frac{ 1 - D^2 A^2 }{ \epsilon D^2 T } },
\end{equation}
where $T$ is the total number of events in the sample $T = N_a^0 + N_b^0$.
The statistical power of different flavor tagging methods varies
as $\epsilon D^2$.

\subsubsection{Measuring the Dilution of the Flavor Tags}

We measure the dilution of the flavor tags from our data sample.
We start by defining a raw dilution, $D_{\rm raw}$:
\begin{equation}
\label{eqn:Draw}
  D_{\rm raw} = \frac{ N_{\rm RS} - N_{\rm WS} }{ N_{\rm RS} + N_{\rm WS} },
\end{equation}
where $N_{\rm RS}$ and $N_{\rm WS}$ are the number of
right-sign and wrong-sign events, respectively.
Right-sign (wrong-sign) means that the charge of the trigger lepton
is opposite to (the same as) the charge of the soft lepton or jet charge.
If the charge of the trigger lepton unambiguously identified
the flavor at production of the $B$~hadron, then $D_{\rm raw}$
would be equal to the true dilution $D$ of the flavor tag.
All wrong-sign events would result from
the flavor tag being incorrect.
However, since some wrong-sign events result from the trigger lepton
coming from a $B$~meson that mixed or from a sequential decay,
$D_{\rm raw}$ is an underestimate of the true dilution of
the flavor tag.
Backgrounds from $c\bar{c}$ production and fake leptons further complicate
the interpretation.
Nevertheless, the true dilution is approximately related to $D_{\rm raw}$
by a scale factor $N_D$:
\begin{equation}
\label{eqn:DrawToD}
  D = N_D \cdot D_{\rm raw}.
\end{equation}
The form of Equation~\ref{eqn:DrawToD} is derived in Appendix~\ref{app:nd}.
We use this estimation of the true dilution to estimate the
probability that the flavor tag is correct on an event-by-event basis:
\begin{equation}
\label{eqn:Ptag}
  P_{\rm tag} = \frac{1}{2}\left( 1 + N_D \cdot D_{\rm raw} \right),
\end{equation}
This probability is used in the measurement of $\Delta m_d$
as described in Section~\ref{sect:fit}.
The dilution normalization $N_D$ is determined simultaneously
with $\Delta m_d$.

\subsubsection{Dilution for the Soft-lepton Tag}

\begin{figure}
\postscript{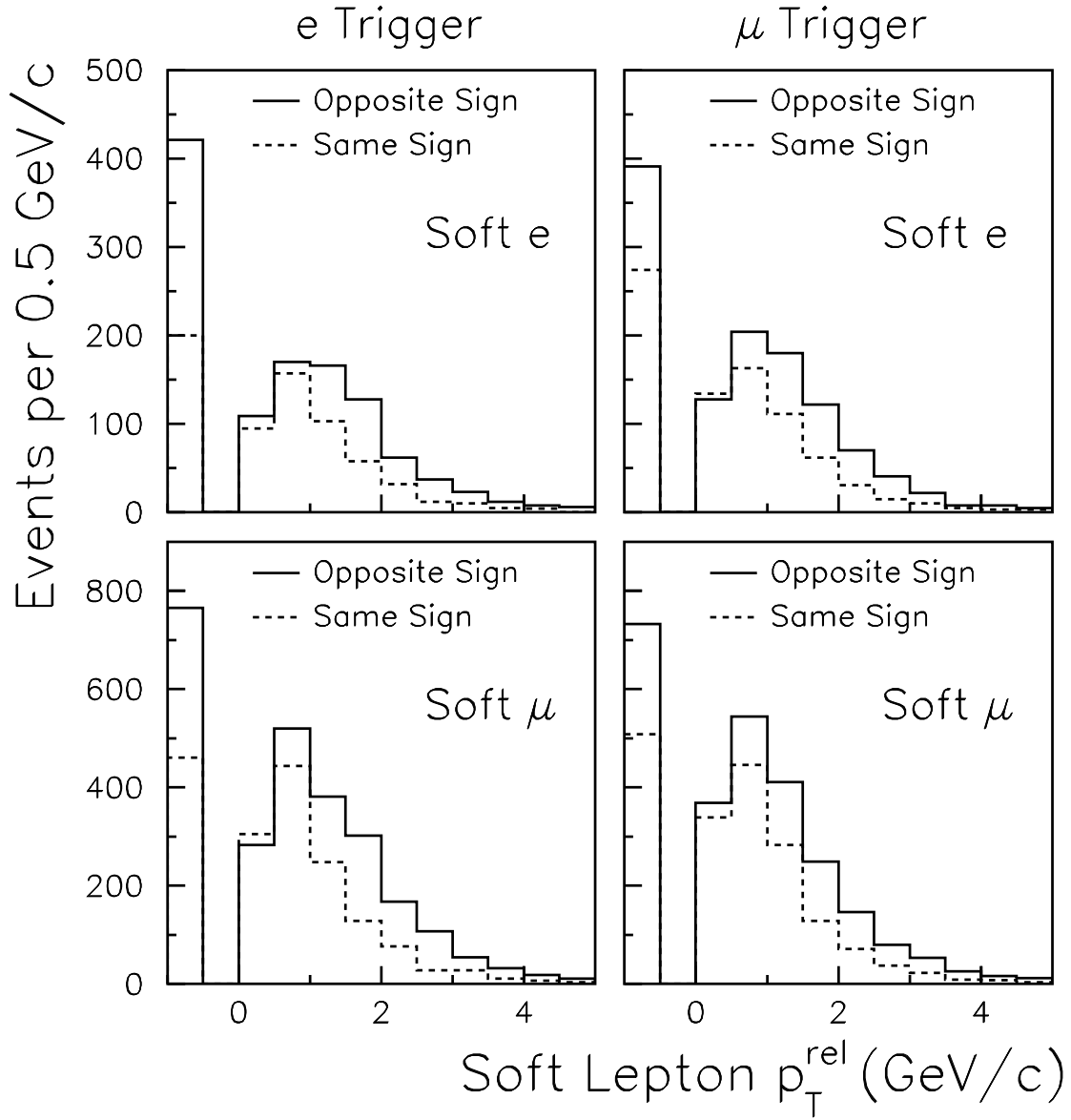}{1.0}
\caption{ 
   Distributions of soft-$e$ and soft-$\mu$ $\ptrel$ for 
   $e$-trigger and $\mu$-trigger events.
   The solid (dashed) distributions are for events where
   the charge of the soft and trigger leptons have opposite (like) sign.
   The bin below 0 is for events where the soft lepton
   was isolated.
}
\label{fig:sltPtrel}
\end{figure}

\begin{figure}
\postscript{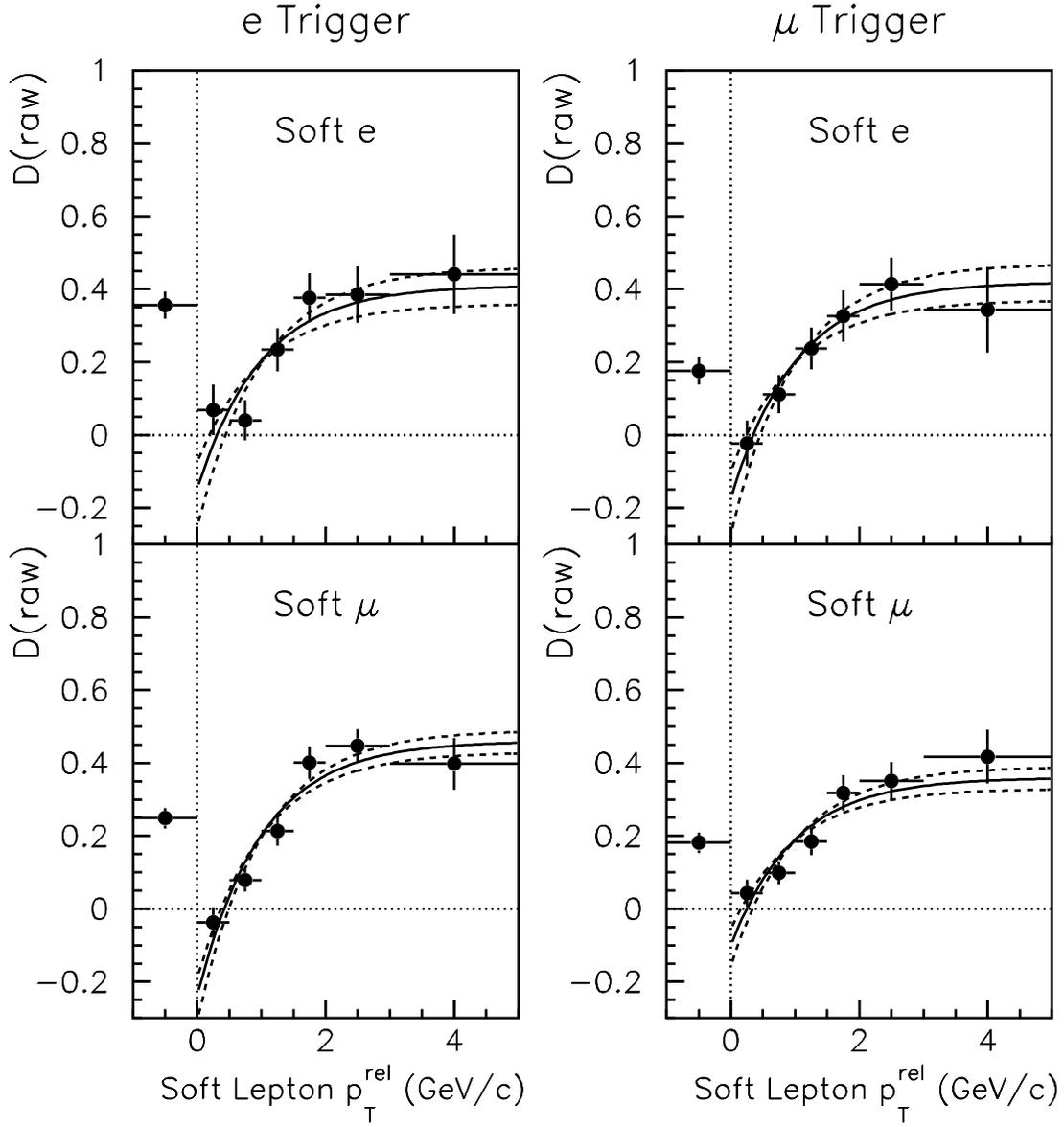}{1.0}
\caption{ 
   Soft-$e$ and soft-$\mu$ tag raw dilution as a function of
   soft-lepton $\ptrel$ for the $e$ and $\mu$ trigger data.
   The solid curve is the parameterization used for
   calculating the expected raw dilution on an
   event-by-event basis based on the soft-lepton $\ptrel$.
   The dashed curves show the variation used in the
   evaluation of the systematic uncertainty on the raw
   dilution parameterization.
}
\label{fig:sltDraw}
\end{figure}

To maximize the effectiveness of the soft-lepton flavor tag,
the data are binned in the $\ptrel$ of the soft lepton.
The $\ptrel$ of the soft lepton is defined in the same way
as the $\ptrel$ of the trigger lepton.
The same-sign and opposite-sign soft-lepton
$\ptrel$ distributions are shown in 
Figure~\ref{fig:sltPtrel},
where the sign comparison is
between the trigger-lepton charge and soft-lepton
charge.  
The $\ptrel<0$ bin is for the case where the soft lepton
is the only track in the jet, so that $\ptrel=0$.
If neither $B$ decayed in a mixed state 
and if both leptons are from semileptonic decay,
the charge of the trigger lepton would be opposite
the charge of the soft lepton.
Figure~\ref{fig:sltDraw} shows the raw dilution
$D_{\rm raw}$ for the soft-lepton tagged events.
The raw dilution is derived from the number of
same-sign and opposite-sign events in each
bin of $\ptrel$.
The dilution is lower for low $\ptrel$ because
fake leptons and leptons from sequential semileptonic
decay tend to have relatively low $\ptrel$ values.
The $\ptrel$ dependence of $D_{\rm raw}$ was parameterized using
the form
\begin{equation}
\label{eqn:sltD}
   D_{\rm raw}(\ptrel) = A \cdot \left( 1 - e^{ - \ptrel + B } \right),
\end{equation}
where $A$ and $B$ are parameters determined from the data.
The average raw dilution is used for isolated soft leptons that
have no measurement of $\ptrel$.
The form of Equation~\ref{eqn:sltD} is empirical and was 
found to describe the shape of $D_{raw}$ as a 
function of $\ptrel$ well in the Monte Carlo.
Parameters $A$ and $B$  are  measured separately 
for the $e$ and $\mu$ trigger data because
the fraction of trigger leptons from sequential decay and
the trigger lepton purity are different for the $e$
and $\mu$ data, which affects the raw dilution.
Parameters $A$ and $B$ are also measured separately
for the soft-$e$ and soft-$\mu$ tags because the
fractions of fake soft leptons may be different
for soft electrons and soft muons.
The fitted values of $A$ and $B$ and the average raw dilution for
isolated (no $\ptrel$) soft leptons for soft-$e$ and soft-$\mu$ tags
in the $e$ and $\mu$ trigger data are listed in Table~\ref{tab:sltAB}.

   The values of $A$ and $B$ determined from the data are assumed
to describe the SLT raw dilution as a function of $\ptrel$ for
$\bbbar$ events.  
This is only an approximation since a small fraction (less than $10\%$)
of the data are $c\bar{c}$ events.
A $c\bar{c}$ event in which $c\rightarrow\ell^+ \nu s$ and
$\bar{c}\rightarrow\ell^- \bar{\nu} \bar{s}$ can produce
opposite-sign events in which one of the leptons is associated
with a secondary vertex, and the other lepton produces a soft-lepton tag.
The soft-lepton tags in $c\bar{c}$ events have a much softer $p_{T}^{\rm rel}$
spectrum than soft-lepton tags in $b\bar{b}$ events.
The Monte Carlo predicts that $\ccbar$ events
affect the values of $A$ and $B$ by an amount less
than the statistical uncertainties on the fitted values
of  $A$ and $B$.
The effects of the approximation above are accounted for
since the values of $A$ and $B$ are varied by their statistical
uncertainties in the determination of the systematic uncertainties
on $\Delta m_d$ and the dilution normalization parameters.

\subsubsection{Dilution for the Jet-charge Tag}

\begin{figure}
\postscript{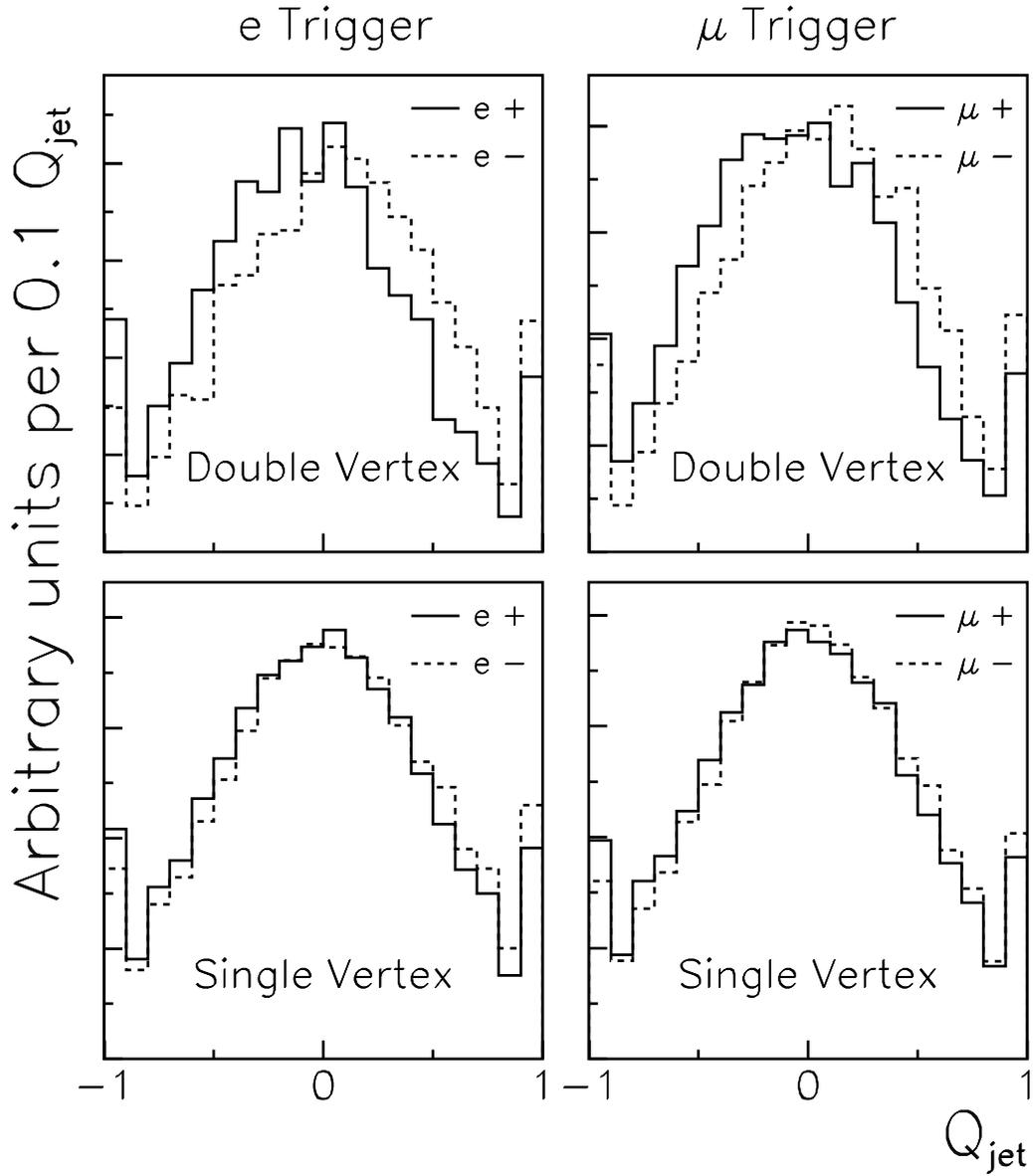}{1.0}
\caption{ 
   Jet charge distributions for double (top) and single
   (bottom) vertex events for the $e$ and $\mu$ trigger data.  
   The data have been divided
   into events with a positively or negatively-charged 
   trigger lepton ($\ell^+$ or $\ell^-$).
   Note that the separation between the $\ell^+$ and 
   $\ell^-$ distributions is significantly larger
   for the double vertex events.
}
\label{fig:JCdist}
\end{figure}

\begin{figure}
\postscript{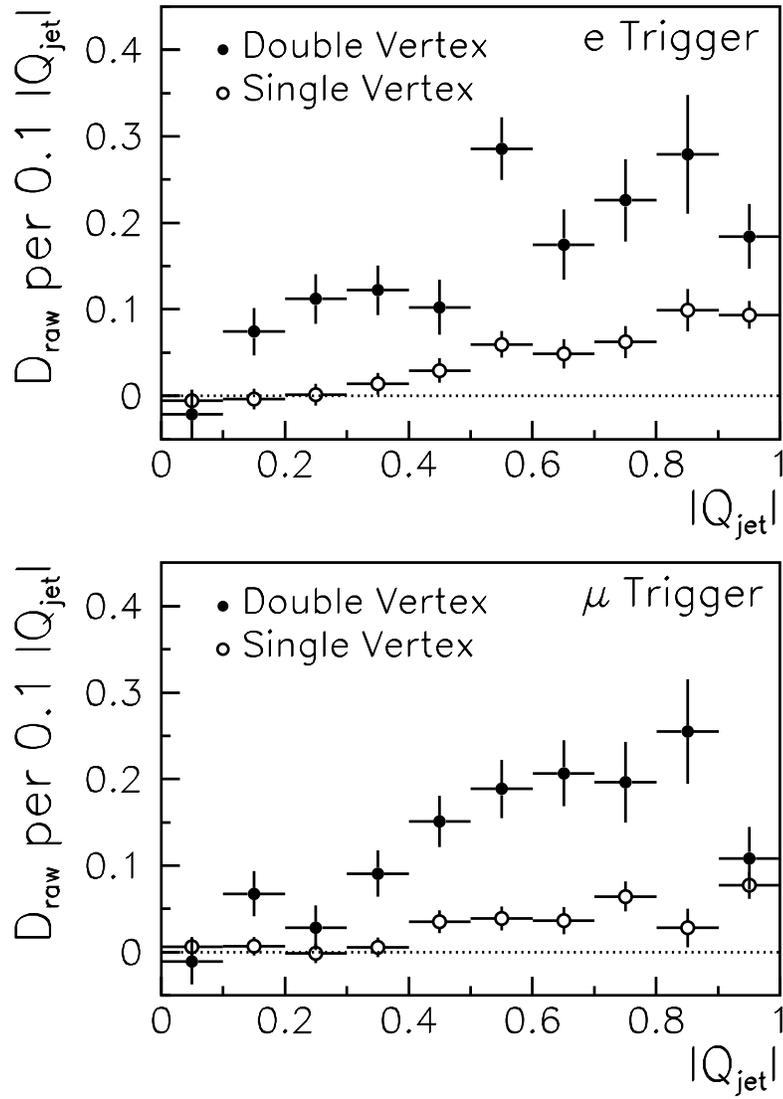}{1.0}
\caption{
   The raw dilution $D_{raw}$ as a function of absolute
   jet charge $|Q_{jet}|$ for double and single vertex
   events in the $e$ and $\mu$ trigger data.  
   Note the significantly higher dilution
   for double vertex events.
}
\label{fig:dplot}
\end{figure}

   Figure~\ref{fig:JCdist} shows the jet charge
distributions for single-vertex and double-vertex events.
The data have been divided into events with
a positively or negatively-charged trigger lepton ($\ell$).
There is an anticorrelation between the sign
of the jet charge and the trigger-lepton charge
on average.
The degree of separation between the $\ell^+$ and
$\ell^-$ distributions is related to the raw dilution
of the jet-charge flavor tag, shown in 
Figure~\ref{fig:dplot} as a function of the
magnitude of the jet charge $|Q_{\rm jet}|$.
For double-vertex events, the presence of the
second secondary vertex increases the probability
that the jet selected for the calculation of
the jet charge is in fact the other $B$ in
the event.
This translates into a significantly higher
raw dilution for double-vertex events.

   The $|Q_{\rm jet}|$ dependence of $D_{\rm raw}$ is used to
predict the probability that the jet-charge tag is
correct on an event-by-event basis, just as $\ptrel$
is used for soft leptons.
The $|Q_{\rm jet}|$ dependence of $D_{\rm raw}$ in the data
in Figure~\ref{fig:dplot} was parameterized with the form
\begin{equation}
\label{eqn:jcD}
  D_{\rm raw}(|Q_{\rm jet}|) = |Q_{\rm jet}| \cdot D_{\rm max}
\end{equation}
excluding events with $|Q_{\rm jet}| = 1$ (the rightmost data point).
For events with $|Q_{\rm jet}| = 1$, the average $|Q_{\rm jet}| = 1$
dilution is used.
The slope $D_{\rm max}$ is determined separately for the
single-vertex and double-vertex events in the $e$ and $\mu$
trigger data, respectively.
These slopes and the average raw dilution for $|Q_{\rm jet}|=1$ 
are listed in Table~\ref{tab:jcDmax}.
We use equation~\ref{eqn:Ptag}, with $D_{\rm raw}$ estimated for each
event using equation~\ref{eqn:sltD}
for SLT events and equation~\ref{eqn:jcD} for JCT events,
in the determination of $\Delta m_d$ to effectively discriminate
between high and low quality flavor tags.


\subsection{Unbinned Maximum Likelihood Fit}

\label{sect:fit}

We use an unbinned maximum likelihood fit to simultaneously
determine $\Delta m_d$ and the dilution normalization $N_D$.
The effectiveness ($\epsilon D^2$) of each flavor tag is derived from the measured
$\epsilon$, $D_{\rm raw}$, and the dilution normalization ($N_D$)
from the fit.
The fraction of $B^0$ mesons that decay in a mixed
state as a function of the proper time at decay
is given by
\begin{equation}
\label{eqn:Fmix}
  F_{\rm mix}(t) = \frac{1}{2}\left( 1 - \cos(\Delta m_d \cdot t) \right).
\end{equation}
In a pure $B_d^0$ sample with perfect flavor tagging and proper 
time resolution, $F_{\rm mix}(t)$ would be equivalent to the fraction
of same-sign events, comparing the sign of the flavors at decay
and production.  
If the flavor tag is imperfect, equation~\ref{eqn:Fmix} becomes
\begin{equation}
\label{eqn:FmixP}
  F_{\rm mix}(t) = P_{\rm tag}    \cdot \frac{1}{2}\left( 1 - \cos(\Delta m_d \cdot t) \right)
             + P_{\rm mistag} \cdot \frac{1}{2}\left( 1 + \cos(\Delta m_d \cdot t) \right),
\end{equation}
where $P_{\rm tag}$ ($P_{\rm mistag}$) is the probability that the flavor tag
is correct (incorrect).
Using the relations
\begin{eqnarray}
\label{eqn:Prtag}
  P_{\rm tag}    & = & \frac{1}{2}( 1 + N_D \cdot D_{\rm raw} ); \\
\label{eqn:Prmistag}
  P_{\rm mistag} & = & \frac{1}{2}( 1 - N_D \cdot D_{\rm raw} );
\end{eqnarray}
equation~\ref{eqn:FmixP} reduces to
\begin{equation}
\label{eqn:FmixPred}
  F_{\rm mix}(t) = \frac{1}{2}\left( 1 - N_D \cdot D_{\rm raw} \cdot \cos(\Delta m_d \cdot t) \right).
\end{equation}
Although $N_D$ and $\Delta m_d$ are correlated,
the basic concept is that $N_D$ is determined from
the amplitude of the oscillation in the same-sign fraction and
$\Delta m_d$ is determined from the frequency.


   We determine $\Delta m_d$ and $N_D$ by minimizing the negative log-likelihood:
\begin{equation}
-\ln{\cal L} = \sum_{i}^{n_{\rm SS}}\ln({\cal P}_{\rm SS}^i) +
               \sum_{j}^{n_{\rm OS}}\ln({\cal P}_{\rm OS}^j),
\end{equation}
where $\cal{P}_{\rm SS}$ is the probability density for
events that are tagged as same-sign, and $\cal{P}_{\rm OS}$
is the probability density for events that are tagged as opposite-sign.
Each event has three inputs into the likelihood:
\begin{enumerate}
\item
the assignment as same-sign or opposite-sign, which is
based on the comparison of the sign of the SLT or JCT flavor tag
with the charge of the trigger lepton,
\item
the estimated probability that the SLT or JCT flavor tag is correct,
using equation~\ref{eqn:Ptag}, with $D_{\rm raw}$ from
equation~\ref{eqn:sltD} for the SLT flavor tag or from
equation~\ref{eqn:jcD} for the JCT flavor tag, and
\item
the decay distance $L_{xy}$.
\end{enumerate}
In addition to the above three inputs, we use $\ptcl$ and $\mcl$ to 
select the $K$~distribution that is used in the determination of
the reconstructed proper decay-distance.
The construction of the probability densities requires several parameters.
These parameters are listed in Table~\ref{tab:fitparam}.

   Both ${\cal P}_{\rm SS}$ and ${\cal P}_{\rm OS}$ are the sum of several terms.
First they have a term for the $\bd$ signal:
\begin{eqnarray}
{\cal P}_{\rm OS}(\bd) = {\cal P}_{\rm tag}{\cal P}_{\rm nomix} +
                         {\cal P}_{\rm mistag}{\cal P}_{\rm mix}; \\
{\cal P}_{\rm SS}(\bd) = {\cal P}_{\rm tag}{\cal P}_{\rm mix} +
                         {\cal P}_{\rm mistag}{\cal P}_{\rm nomix};
\end{eqnarray}
where ${\cal P}_{\rm tag}$ and ${\cal P}_{\rm mistag}$ are given by
equations~\ref{eqn:Prtag} and~\ref{eqn:Prmistag}, respectively,
and ${\cal P}_{\rm nomix}$ and ${\cal P}_{\rm mix}$ are given
by equations~\ref{eqn:pnomix} and~\ref{eqn:pmixed}, respectively.
Next there are terms for the other $B$~hadrons, including terms for
both $\bplus$ and $b$~baryons, as well as a term for $\bs$.
The terms for $\bplus$ are
\begin{eqnarray}
{\cal P}_{\rm OS}(\bplus) =
 {\cal P}_{\rm tag} \frac{1}{\tau_{\bplus}} e^{-t/\tau_{\bplus}};
                                                                \\
 {\cal P}_{\rm SS}(\bplus) =
 {\cal P}_{\rm mistag} \frac{1}{\tau_{\bplus}} e^{-t/\tau_{\bplus}};
\end{eqnarray}
where $\tau_{\bplus}$ is the lifetime of the $\bplus$, and $t$ is the
proper decay-time.
The terms for $b$~baryons are similar, except that
$\tau_{\bplus}$ is replaced by the lifetime of the
$b$~baryons, $\tau_{\rm baryon}$.
The terms for $\bs$ are similar to the terms for $\bd$, except that
$\Delta m_d$ is replaced by $\Delta m_s$, which is assumed to be
very large ({\em i.e.}~beyond our experimental sensitivity)
so that these terms effectively look like
\begin{equation}
{\cal P}_{\rm OS}(\bs) = {\cal P}_{\rm SS}(\bs) =
 \frac{1}{2 \tau_{\bs}} e^{-t/\tau_{\bs}}.
\end{equation}
The values of the lifetimes of the $B$~hadrons and the value of $\Delta m_s$
used in the probability densities are listed in Table~\ref{tab:fitparam}.
Each term for the various $B$~hadrons is multiplied by the expected
relative contribution,
$f_{\bd}$, $f_{\bplus}$, $f_{\bs}$, and $f_{\rm baryon}$,
of these various hadrons to the data sample.
The production fractions are renormalized to take into account the
different semileptonic branching fractions of the various $B$~hadrons:
the semileptonic widths are assumed to be identical, so the 
semileptonic branching fractions are scaled to agree with the
relative lifetimes.

   In addition to the terms for direct semileptonic decay, there are terms
that take into account the contribution from sequential decays.
The fraction of sequential decays $f_{\rm seq}$ is listed in
Table~\ref{tab:fitparam}; 75\% of these sequential decays produce
leptons with a sign opposite to direct semileptonic decay.

   The terms that take into account the contribution from $\ccbar$ events
are similar to the terms for $\bplus$, except that we use
an assumed flavor tagging dilution
and an effective lifetime $\tau_{\ccbar}$.
The true flavor tagging dilution is not known {\it a priori} for 
both bottom and charm decays.
Just as for $B$ decays, 
we do not consider the Monte Carlo reliable for predicting the
JCT or SLT dilution for charm decays, so we use assumed 
values for the charm dilution and varied them by the maximum
possible amount for the systematic uncertainties.
For the JCT, we expect the dilution for charm decays
to be worse than for bottom decays,
therefore we assume
$D_{\ccbar}/D_{\bbbar} = 0.5$ for the JCT,
and vary $D_{\ccbar}/D_{\bbbar}$ from 0 to 1 in the evaluation
of the systematic errors.
The ratio $D_{\ccbar}/D_{\bbbar}$ is used to rescale the 
predicted $\bbbar$ dilution for the event,  based
on $|Q_{jet}|$ using Equation~\ref{eqn:jcD},
to give the predicted $\ccbar$ dilution for the event.
The SLT dilution for charm decays could be anything from
0 to 1 depending on the fraction of fake soft leptons in $\ccbar$
events. 
Unlike bottom decays, the SLT dilution for charm 
decays does not fall near
$\ptrel=0$ due to soft leptons from sequential decays, therefore we assume
$D_{\ccbar} = 0.5$,
independent of $\ptrel$
for the SLT,
and vary $D_{\ccbar}/D_{\bbbar}$ from 0 to 1 in the evaluation
of the systematic errors.

The effective lifetime $\tau_{\ccbar}=1.53$~ps
is determined from the $c\bar{c}$ Monte Carlo samples,
where the proper time at decay was reconstructed using the $K$~distributions 
from the $b\bar{b}$ Monte Carlo and equation~\ref{eqn:ctau}.
The relative fractions of $\bbbar$ ($F_{\bbbar}$) and $\ccbar$ ($F_{\ccbar}$),
which can be determined from Table~\ref{tab:mclptrel},
multiply the terms in the likelihood corresponding to
$\bbbar$ and $\ccbar$ production, respectively.
Finally, for the case of the $\mu$-trigger data,
a term for fake muons was included with 
the $B^0$ lifetime and zero dilution.
The relative fraction of fake muons $F^{\: 0}_{{\rm fake}\ \mu}$
(discussed in Appendix~\ref{app:fake_muons}) is listed
in Table~\ref{tab:fitparam}.
In this case, $F_{\bbbar}$ and $F_{\ccbar}$ are scaled by
$1-F^{\: 0}_{{\rm fake}\ \mu}$.

\begin{figure}
\postscript{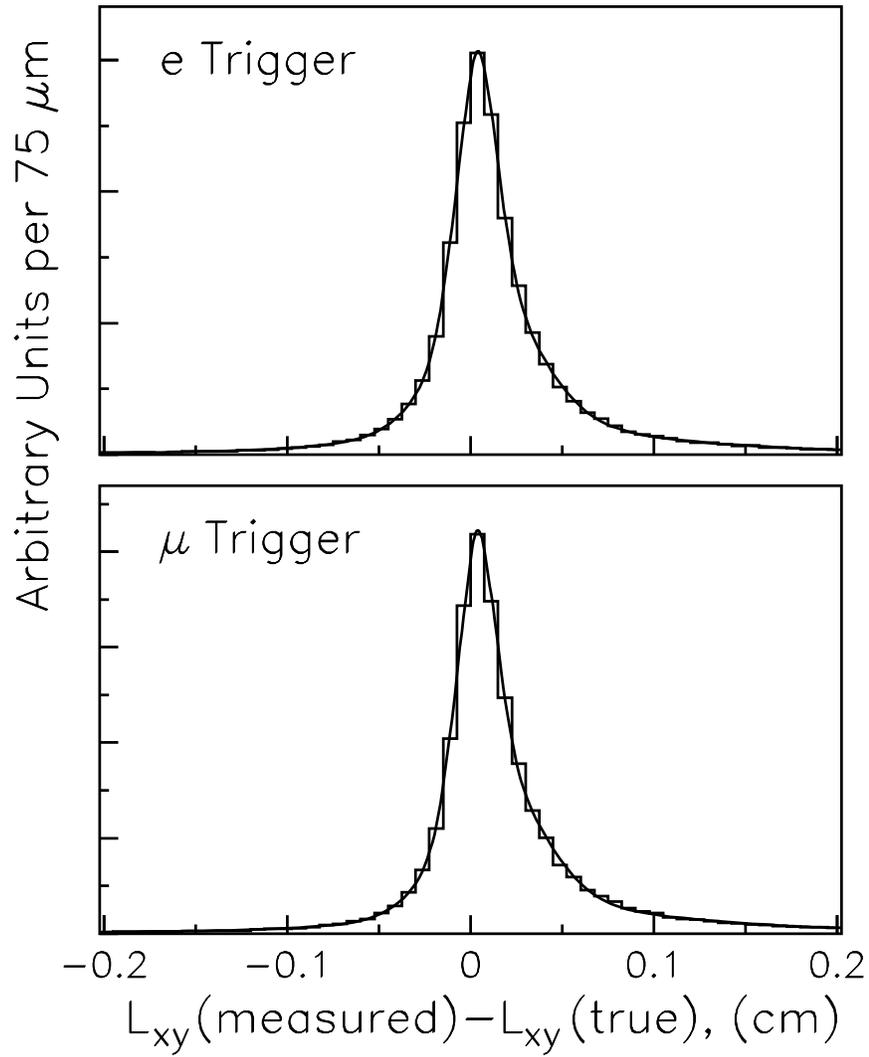}{1.0}
\caption{
   The transverse decay length resolution functions,
defined as the difference between the measured and
true $L_{xy}$ in the $b\bar{b}$ Monte Carlo, for
the $e$ and $\mu$ trigger  data.
The distributions were
fit to three Gaussians and two exponentials.
}
\label{fig:LxyRes}
\end{figure}

   The probability densities $P_{\rm SS}(t)$ and $P_{\rm OS}(t)$ are functions
of the true proper time at decay $t$.  
We take into account the experimental resolution on $t$ by convoluting
$P_{\rm SS}(t)$ and $P_{\rm OS}(t)$ with $L_{xy}$ and $\ptbprime$
resolution functions.
Distributions of 
$\delta L_{xy} \equiv L_{xy}({\rm measured}) - L_{xy}({\rm true})$
from the $e$ and $\mu$ trigger $b\bar{b}$ Monte Carlo samples are
parameterized with the sum of three Gaussians,
a $\delta L_{xy}>0$ exponential, and a $\delta L_{xy}<0$ exponential.
The $\delta L_{xy}$ distributions and their parameterizations
are shown in Figure~\ref{fig:LxyRes} 
for the $e$ and $\mu$ trigger data.
The $K$~distributions, like those shown in Figure~\ref{fig:Kfac},
are used as $\ptbprime$ resolution functions.
The $L_{xy}$ and $\ptbprime$ resolution functions 
describe the $t$ smearing for $b\bar{b}$
events from direct production ($gg \rightarrow b\bar{b}$ and
$q\bar{q} \rightarrow b\bar{b}$).  
They do not describe backgrounds such as photon conversion events
(for the $e$ trigger) or events with a fake trigger muon.
They also do not describe $b\bar{b}$ events from gluon splitting
($gg\rightarrow gg$ followed by $g\rightarrow b\bar{b}$), which
tend to have worse $t$ resolution due to secondary vertices that
include decay products from both $B$~hadrons.
While the total amounts of these backgrounds are reasonably small,
they become more important near $ct=0$ and for very large $ct$
values (beyond several $B$ lifetimes).
For these reasons, we only use events with
$0.02 \ {\rm cm} < ct < 0.30 \ {\rm cm}$
in the fit for $\Delta m_d$ and the dilution normalization $N_D$.


\section{Fit Results}

    The free parameters in the unbinned maximum likelihood fit are
the mass difference $\Delta m_d$ and the dilution normalization
factors $N_D$.
There are six dilution normalization factors:
\begin{enumerate}
\item
$N_{D,{\rm JCSV}}^e$ and $N_{D,{\rm JCSV}}^\mu$:
the dilution normalization factors for the jet-charge flavor tag
in the case that the jet does not contain a secondary vertex
for the electron-trigger and muon-trigger data samples, respectively.
\item
$N_{D,{\rm JCDV}}^e$ and $N_{D,{\rm JCDV}}^\mu$:
the dilution normalization factors for the jet-charge flavor tag
in the case that the jet does contain a secondary vertex
for the electron-trigger and muon-trigger data samples, respectively.
\item
$N_{D,{\rm SLT}}^e$ and $N_{D,{\rm SLT}}^\mu$:
the dilution normalization factors for the soft-lepton flavor tag
for the electron-trigger and muon-trigger data samples, respectively.
\end{enumerate}
To determine the dilution normalization factors needed to calculate
the flavor tag $\epsilon D^2$ values (see Section~\ref{sec:eds}),
the data are grouped into four subsamples:
\begin{enumerate}
\item $e$-trigger, soft-lepton flavor tag,
\item $e$-trigger, jet-charge flavor tag,
\item $\mu$-trigger, soft-lepton flavor tag, and 
\item $\mu$-trigger, jet-charge flavor tag.
\end{enumerate}
There is some overlap of events in these four subsamples as some
events have both a SLT and a JCT.
About 20\% of the events with soft-lepton tags
did not pass the single-lepton Level~2 trigger
and came, instead, from a dilepton trigger.
In order for the SLT efficiency to be well defined, we require
that the SLT events pass the single-lepton trigger in data
subsamples 1 and 3, which are used to
determine $N^e_{D,{\rm SLT}}$, $N^\mu_{D,{\rm SLT}}$, and the
SLT efficiencies needed for calculating the SLT $\epsilon D^2$.

   The fit results for the individual flavor taggers are given
in Table~\ref{tab:indivFits}.
The dilution normalization factors are given for two cases:
$\Delta m_d$ free to float in the fit and $\Delta m_d$ fixed
to the world average~\cite{PDG} ( 0.474 $\hbar \ {\rm ps}^{-1}$).
The value of $\Delta m_d$ is held fixed to the world average because 
$\Delta m_d$ and the $N_D$ factors are correlated.
The correlation coefficients between $\Delta m_d$ and the
$N_D$ constants range from 0.55 to 0.81.
Fixing $\Delta m_d$ reduces the statistical uncertainty on 
the dilution normalization constants and removes any 
bias from statistical fluctuations that pull
$\Delta m_d$ high or low.
The $N_D$ factors determined with $\Delta m_d$ fixed to the world
average are used in the calculation of the flavor tag $\epsilon D^2$ values.
The $N_D$ factors in Table~\ref{tab:indivFits} 
are consistent with our expectations from the composition of
the data (see Appendix~\ref{app:nd}).

To determine $\Delta m_d$, we fit all four data subsamples simultaneously.
For events with both a SLT and a JCT, we use the SLT because
it has significantly higher average dilution.
The results of the simultaneous fit of 
the $e$ and $\mu$ trigger data using both flavor tagging methods
are listed in Table~\ref{tab:fit_results}.
We find $\Delta m_d = 0.500 \pm 0.052 \ \hbar \ {\rm ps}^{-1}$,
where the uncertainty is statistical only.
This is consistent with the world average value 
(0.464 $\pm$ 0.018 $\hbar \ {\rm ps}^{-1}$)~\cite{PDG98}.
Using the SLT tag in doubly flavor tagged events removes
events with higher-than-average dilution from the JCT.
This results in lower values of $N_D$ for the jet-charge flavor tag
than found in the fits of the individual samples listed
in Table~\ref{tab:indivFits}.

\begin{figure}
\postscript{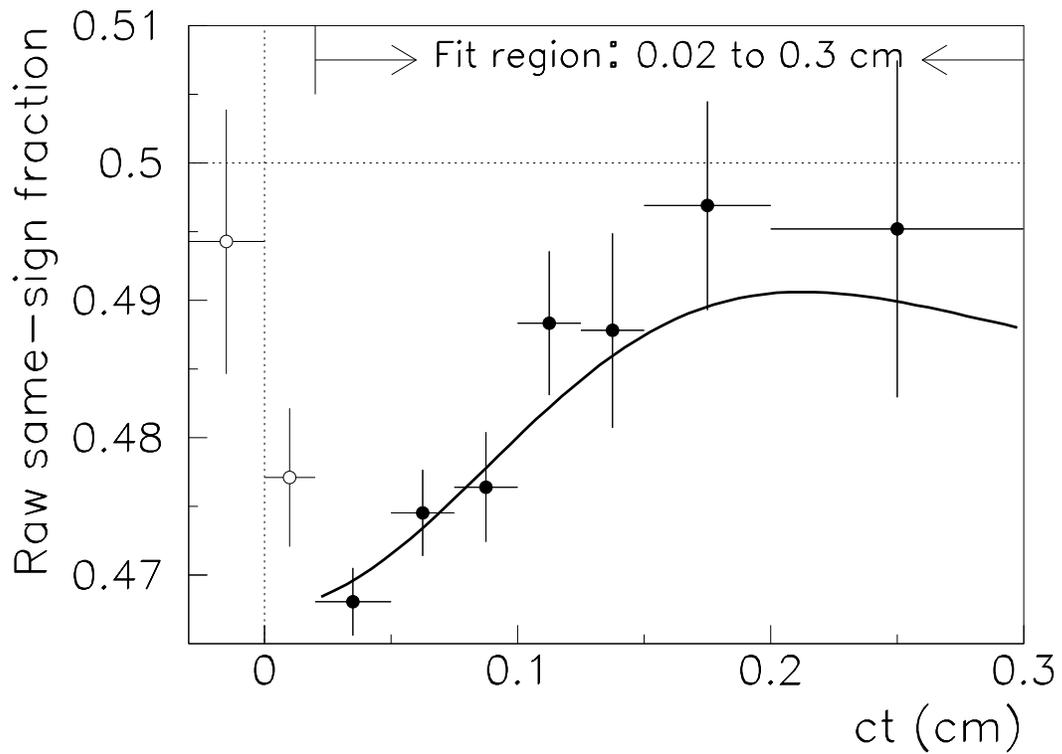}{1.0}
\caption{ 
   The same sign fraction as a function of the proper
   decay length $ct$.  A representation of the 
   unbinned maximum likelihood fit for $\Delta m_d$
   and the $N_D$ factors is superimposed on the data.  
}
\label{fig:ssfr}
\end{figure}

   Figure~\ref{fig:ssfr} shows the fraction of same-sign events as
a function of the reconstructed proper decay length $ct$.
The points with error bars are the data.
The curve is a representation of the results of the fit for
$\Delta m_d$ and $N_D$.
Figure~\ref{fig:ssfr} has been included to illustrate the
clear evidence of $B_d^0$ mixing in the data.
However, it does not contain all of the information that goes
into the unbinned likelihood fit.
In Figure~\ref{fig:ssfr}, all events are treated equally. 
In the fit, events are effectively weighted based on their
estimated dilution.


\subsection{ Check of Fitting Procedure }

\begin{figure}
\postscript{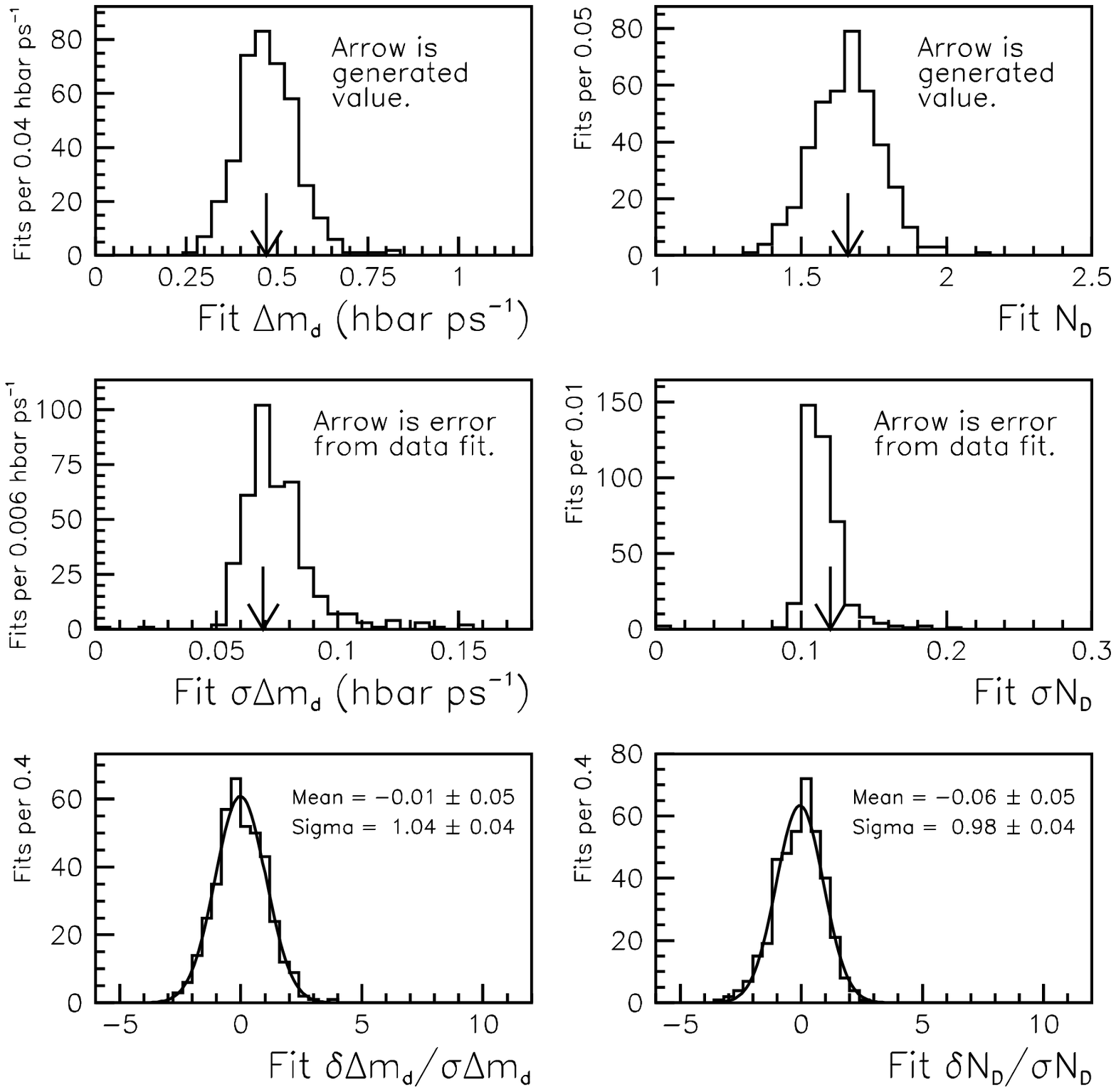}{1.0}
\caption{
  Fit results for 400 fast Monte Carlo samples, simulating
the $e$-trigger, SLT flavor tagged data.  
The top row of plots show distributions of
the fitted values with arrows indicating 
the generated values for the samples.
The middle plots show distributions of
the statistical uncertainties on the
fitted values with arrows indicating the statistical
uncertainty for the fit to the real data.
The bottom plots show distributions of the deviation 
from the true value, divided by the statistical uncertainty
(pull distributions).
The pull distributions have zero mean and unit width,
as expected.
}
\label{fig:toy}
\end{figure}

   To check our fitting procedure, we used a
fast Monte Carlo which generated hundreds of 
data samples, each with the same statistics,
tagging dilution, and $t$ resolution as the real data.
Figure~\ref{fig:toy} shows the fit results of 400 
fast Monte Carlo samples, representing the SLT
flavor tagged, $e$-trigger data.
The top row of plots show distributions of the
fitted values of $\Delta m_d$ and $N_D$ for the
400 samples.  
The arrows indicate the values of $\Delta m_d$
and $N_D$ with which the samples were generated.
The mean of each distribution is consistent with
the generation value.
The middle row of plots show the distributions
of the statistical uncertainty on $\Delta m_d$ and
$N_D$.  
The arrows indicate the statistical uncertainty from the
fit to the SLT flavor tagged,  $e$-trigger data.
The statistical uncertainties on $\Delta m_d$ and 
$N_D$ for the data are near the most probable values from
the fast Monte Carlo samples.
The bottom row of plots show the distributions
of the deviation of the fitted value from the
generation value divided by the statistical uncertainty
for $\Delta m_d$ and $N_D$.
These distributions have a mean of zero and unit width,
which confirms that $\Delta m_d$ and $N_D$ are
unbiased and that the statistical
uncertainty is correct.


\subsection{ Systematic Uncertainties }

To determine the systematic uncertainty on $\Delta m_d$,
the fixed input parameters in Table~\ref{tab:fitparam}
were varied by the amounts listed in this Table.
To determine the systematic uncertainty associated with the
dilution parameterizations,
the parameters describing the $D_{\rm raw}$ dependence on
$|Q_{\rm jet}|$ and $\ptrel$ of the soft lepton were
varied by their statistical uncertainties.
A systematic uncertainty to account for the possible presence of
fake secondary vertices from non-heavy flavor backgrounds as well as
heavy flavor events (gluon splitting) 
was determined using a combination
of data and Monte Carlo.
The observed excess in the data over the combined contribution
of $\bbbar$ and $\ccbar$ (see Figure~\ref{fig:ctau})
was used to define the shape of this background.
This shape was included in the likelihood function with a
dilution that varied from 0 to $D_{b\bar{b}}$.

The systematic uncertainty assigned to the variation of each input
parameter was the shift in $\Delta m_d$ from the fit result
with the nominal values of the input parameters.
The total systematic uncertainty is the sum in quadrature of
these shifts.
The same procedure was applied to the four data subsamples
($e$ SLT, $e$ JCT, $\mu$ SLT, $\mu$ JCT) to determine
the systematic uncertainties for the dilution normalization
factors.

The procedure described above was checked 
for the largest individual contributions to the
systematic uncertainty using fast Monte Carlo samples.  
Samples generated with a variation on
one of the input parameters (e.g. $\tau_{B^+}/\tau_{B^0} = 1.02 + 0.05$)
were fit using the nominal parameter (e.g. $\tau_{B^+}/\tau_{B^0} = 1.02$).
The average bias on the fitted values for the
fast Monte Carlo samples was consistent
with the deviation on the fitted values 
observed when the data are fit with a fixed parameter 
variation.

Table~\ref{tab:dmdsys} lists the individual contributions
to the systematic uncertainty on $\Delta m_d$ ($\pm 0.043 \ \hbar \ {\rm ps}^{-1}$).
The largest single contribution ($\pm 0.032 \ \hbar \ {\rm ps}^{-1}$) 
is the unknown soft-lepton flavor tag dilution
for $\ccbar$ events.
The SLT $\ccbar$ dilution was varied over the
full possible range (0 to 1).
If the assumed $\ccbar$ dilution in the fit is lower (higher) than
its true value, the fraction of same-sign events at small $t$ will appear
lower (higher), which biases 
$\Delta m_d$ low (high).
The second largest contribution ($\pm 0.021 \ \hbar \ {\rm ps}^{-1}$)
is the uncertainty on the
lifetime ratio of the $B^+$ and the $B^0$ mesons.
If the value of $\tau_{B^+}/\tau_{B^0}$ in the fit is
higher (lower) than its true value, the fraction of same-sign
events at short $t$ values will appear larger (smaller), 
which biases $\Delta m_d$ high (low).
The third largest contribution ($\pm 0.009 \ \hbar \ {\rm ps}^{-1}$)
is the uncertainty on the SLT raw dilution parameterization:
$D_{raw}$ as a function of soft-lepton $\ptrel$.
The size of the variation for the SLT is shown by the dashed
curves in Figure~\ref{fig:sltDraw}.
The raw dilution for events with no $\ptrel$ measurement
was independently varied by the statistical uncertainty on the raw dilution
for no-$\ptrel$ events.

Tables~\ref{tab:ndsys_e} and~\ref{tab:ndsys_m} list the
contributions to the systematic uncertainty on the dilution
normalization factors for the $e$ and $\mu$ trigger data
respectively.
As for the systematic uncertainty on $\Delta m_d$, the largest 
contribution comes from the unknown dilution for $\ccbar$ events.
The fraction of $B_s$ events, which we assume are half same-sign
and half-opposite sign, and the fraction of events where
the trigger lepton is from a sequential decay both 
affect the assignment of same-sign events.
If these fractions are low (high), more (less) same-sign events will
be attributed to mistags, thus they are strongly coupled
to the dilution normalization.
The uncertainty on the raw dilution parameterizations also
has a large effect on the dilution normalization systematic
uncertainty.


\subsection{ Flavor Tag $\epsilon D^2$ }
\label{sec:eds}

The measurement of the statistical power $\epsilon D^2$ of the flavor
tagging methods is done in two steps.  
First, the raw $\epsilon D^2$ is calculated using
the raw dilution rather than the true dilution.
Then, the raw $\epsilon D^2$ is rescaled by 
$N_D^2$, which translates the raw dilution to
the true dilution.
The $N_D$ factor for each flavor tagging method
in the $e$ and $\mu$ samples is determined 
in the unbinned maximum likelihood fit
with $\Delta m_d$ fixed to the world
average~\cite{PDG} (0.474 $\hbar \ {\rm ps}^{-1}$).
These $N_D$ values are given in Table~\ref{tab:indivFits}, 
where the first uncertainty is statistical
and the second systematic.

   The raw or uncorrected value of $\epsilon D^2$ is obtained
by summing the $\epsilon D_{\rm raw}^2$ values in the bins
of either $\ptrel$ for the SLT or $|Q_{\rm jet}|$ for the JCT.
The efficiency for
each bin is defined as the number of events in the
bin divided by the total number of events before
flavor tagging.
Table~\ref{tab:eds} lists the total efficiency, 
raw $\epsilon D^2$, dilution normalization,
and true $\epsilon D^2$ for each of the flavor
tagging methods.
Taking the average of the $e$ and $\mu$ trigger data,
we find $\epsilon D^2$ to be ($0.78 \pm 0.12 \pm 0.08$)\%
for the JCT and ($0.91 \pm 0.10 \pm 0.11$)\% for the
SLT where the first uncertainty is statistical and the 
second systematic.
These $\epsilon D^2$ values are about one order of
magnitude lower than typical flavor tagging techniques
employed on the $Z^0$ resonance~\cite{JetCharge}, 
however the large $\bbbar$ cross section in $p\bar{p}$
collisions at $\sqrt{s}=1.8$~TeV yields $\bbbar$ samples
that are about one order of magnitude larger at CDF than 
those collected on the $Z^0$ resonance.
The statistical uncertainty for our measurement of $\Delta m_d$
is still competitive with similar measurements 
on the $Z^0$
resonance, since our smaller $\epsilon D^2$ is compensated
by our larger sample size.

The values of $\epsilon D^2$ for these flavor tags depend
on the data sample in which they are used.
In particular, during next run of the Tevatron, we will collect large samples
of $B^0/\bar{B}^0\rightarrow J/\psi K^0_{\rm S}$
(for the precise measurement of the $CP$~asymmetry parameter $\sin(2\beta)$)
and hadronic $B^0_s$ decays (for the precise determination of $\Delta m_s$).
The triggers used to collect these data samples will be different from
the inclusive lepton trigger used to collect the data for this analysis.
As a result, the $B$~hadron production properties ({\it e.g.}, $p_T$ of the
$B$) are different, and this affects $\epsilon D^2$.
Despite these differences, the results in this paper demonstrate that both
the jet-charge and the soft-lepton flavor tagging methods are viable in the
environment of $p\bar{p}$ collisions.

\section{Summary}

   We have measured $\Delta m_d$ using soft-lepton
and jet-charge flavor tagging methods.  
This is the first application of jet-charge flavor
tagging in a hadron-collider environment.
The flavor at decay was inferred from the charge of
the trigger lepton, which was assumed to be the
product of semileptonic $B$ decay.
The initial flavor was inferred from the other
$B$ in the event, either using the charge of a
soft lepton or the jet charge of the other $B$.
The proper-time at decay for each event was determined from
a partial reconstruction of the decay vertex of the
$B$ that produced the trigger lepton and an estimate
of the $B$ momentum.
The value of $\Delta m_d$ was determined with 
an unbinned maximum likelihood fit 
of the same-sign and opposite-sign proper-time
distributions (comparing
the sign of the trigger-lepton charge and
the flavor tag).
The statistical power of the flavor tagging methods
was measured in the unbinned maximum likelihood
fit by fitting for a scale factor $N_D$, for each
of the flavor tagging methods, which is the ratio of the
raw dilution and the true dilution.

We find 
  $\Delta m_d = 0.500 \pm 0.052 \pm 0.043 \ \hbar \ {\rm ps}^{-1}$,
where the first uncertainty is statistical and the second systematic.
This is consistent with the world average value of 
  $0.464 \pm 0.018 \ \hbar \ {\rm ps}^{-1}$~\cite{PDG98}
and competitive in precision with other
individual measurements of $\Delta m_d$.
We quantify the statistical power of the flavor tagging methods
with  $\epsilon D^2$, which is the tagging efficiency multiplied
by the square of the dilution. 
We find $\epsilon D^2$ to be ($0.78 \pm 0.12 \pm 0.08$)\% for the
jet-charge flavor tag and ($0.91 \pm 0.10 \pm 0.11$)\% for the soft-lepton 
flavor tag, where the first uncertainty is statistical and the second
systematic.
These $\epsilon D^2$ are much lower than what has been achieved
in experiments on the $Z^0$ resonance, however we have demonstrated
that the much higher $\bbbar$ cross-section at the Tevatron
($p\bar{p}$, $\sqrt{s}=1.8$ TeV) can be used to compensate for the
disadvantage in $\epsilon D^2$. 
The jet-charge and soft-lepton flavor tagging techniques
will be important tools in the study of $CP$ violation in the up-coming 
run of the Tevatron.


\section*{Acknowledgments}

We thank the Fermilab staff and the technical staffs of the
participating institutions for their vital contributions.  This work was
supported by the U.S. Department of Energy and National Science Foundation;
the Italian Istituto Nazionale di Fisica Nucleare; the Ministry of Education,
Science and Culture of Japan; the Natural Sciences and Engineering Research
Council of Canada; the National Science Council of the Republic of China;
the Swiss National Science Foundation; and the A. P. Sloan Foundation.

\appendix

\newpage


\section{ The Dilution Normalization Factor }
\label{app:nd}

The true dilution $D$ of a flavor tagging method is defined as
\begin{equation}
   D = 2 \cdot P_{\rm tag} - 1,
\end{equation}
where $P_{\rm tag}$ is the probability that the flavor tag
is correct.
An equivalent expression for $D$ is
\begin{equation}
\label{eqn:appd}
   D = \frac{ N_T - N_M }{ N_T + N_M },
\end{equation}
where $N_T$ ($N_M$) is the number of correct (incorrect)
tags in a sample of $N_{\rm total} = N_T + N_M$ events.
The raw dilution is defined as
\begin{equation}
\label{eqn:appdraw}
  D_{\rm raw}  = \frac{ N_{OS} - N_{SS} }{ N_{OS} + N_{SS} },
\end{equation}
where $N_{OS}$ ($N_{SS}$) is the number of opposite-sign
(same-sign) events in the sample, comparing the trigger-lepton
charge with either the sign of the soft-lepton charge or
the jet charge.
If the data were pure $\bbbar$ with no $B$ mixing and all of
the trigger leptons were from direct $B$ decay, 
all opposite (same) sign events would be correctly (incorrectly)
flavor-tagged.
That is, we would have $N_{T}=N_{OS}$, $N_{M}=N_{SS}$, and
 $D = D_{\rm raw}$.
There are, however, several things in the data that break the
$N_{T}=N_{OS}$ and $N_{M}=N_{SS}$ assumptions.
They are
\begin{itemize}
\item {\bf $B$ mixing}:  If the trigger lepton is from a $B$ hadron
   that decays in a state opposite its original flavor, the 
   trigger-lepton charge will have the ``wrong'' sign. 
   In this case, events with the correct flavor tag are same-sign.
\item {\bf Sequential decays}:  The charge of trigger leptons from
   sequential $B$ decay (\mbox{$b\rightarrow c\rightarrow \ell \ s \ X$})
   is opposite that of direct $B$ decay.  
   For trigger leptons that are from sequential decay,
   events with the correct flavor tag are same-sign,
   if the trigger-lepton $B$ did not mix. 
\item {\bf $\ccbar$ events}:  Events from $\ccbar$ production 
   may have a non-zero dilution that is not the same as
   the dilution from $\bbbar$ events.  
\item {\bf Fake leptons}:  The $e$-trigger data has essentially
   no fake trigger electrons.  However, about 12\% of the 
   $\mu$-trigger data have a hadron that faked a muon, whose
   charge is random (see Appendix~\ref{app:fake_muons}).
\end{itemize}

   If there were no fake leptons,
the number of opposite-sign and
same-sign events from $\bbbar$ production are given by
\begin{eqnarray}
  \label{eqn:nosbb}
   N^{\bbbar}_{OS} & = & (1 - f^{\rm ws}_{\rm seq}) 
               \left[ ( 1  - \bar{\chi}' ) N^{\bbbar}_T + \bar{\chi}' N^{\bbbar}_M \right]
              \nonumber \\
  & & + f^{\rm ws}_{\rm seq} 
               \left[ \bar{\chi}' N^{\bbbar}_T + ( 1  - \bar{\chi}' ) N^{\bbbar}_M \right] \\
  \label{eqn:nssbb}
   N^{\bbbar}_{SS} & = & (1 - f^{\rm ws}_{\rm seq}) 
               \left[ \bar{\chi}'  N^{\bbbar}_T + ( 1 - \bar{\chi}') N^{\bbbar}_M \right]
              \nonumber \\
  & & + f^{\rm ws}_{\rm seq} 
               \left[ ( 1 - \bar{\chi}' ) N^{\bbbar}_T +  \bar{\chi}'  N^{\bbbar}_M \right] 
\end{eqnarray}
where $f^{\rm ws}_{\rm seq}$ is the fraction of trigger leptons in $\bbbar$ events
that are from sequential decay in which the trigger-lepton charge has the ``wrong'' sign,
$\bar{\chi}'$ is the 
effective\footnote{ It is an ``effective'' probability 
  because our  secondary vertexing method is inefficient for low
  values of $t$, which causes $\bar{\chi}'$ to be larger than
  $\bar{\chi}$~\protect{\cite{PDG98}}.}
probability that the $B$ hadron that produced the
trigger lepton decayed in a mixed state, and $N^{\bbbar}_{OS}$ ($N^{\bbbar}_{SS}$)
is the number of same-sign (opposite-sign) $\bbbar$ events.
For events from $\ccbar$ production we have
\begin{eqnarray}
  \label{eqn:noscc}
   N^{\ccbar}_{OS} & = & N^{\ccbar}_T \\
  \label{eqn:nsscc}
   N^{\ccbar}_{SS} & = & N^{\ccbar}_M 
\end{eqnarray}
where $N^{\ccbar}_{OS}$ ($N^{\ccbar}_{SS}$)
is the number of same-sign (opposite-sign) $\ccbar$ events.
Using Equations~\ref{eqn:appdraw}, \ref{eqn:nosbb}, \ref{eqn:nssbb}, \ref{eqn:noscc},
and~\ref{eqn:nsscc} the raw dilution can be written as
\begin{equation}
\label{eqn:draw1}
D_{\rm raw} = \frac{ (1 - 2 \bar{\chi}')(1 - 2 f^{\rm ws}_{\rm seq}) 
          (N^{\bbbar}_T - N^{\bbbar}_M)  +  (N^{\ccbar}_T - N^{\ccbar}_M) }
   {  N^{\bbbar}_T + N^{\bbbar}_M   +   N^{\ccbar}_T + N^{\ccbar}_M  }
\end{equation}

    If a fraction of the events ($F^{\: 0}_{{\rm fake} \ \ell}$) have a fake trigger lepton 
whose charge-sign is random, these events will have a raw dilution of zero since the number of
same-sign fake-lepton events will equal the number of opposite-sign fake-lepton events.
Taking fake leptons into account gives
\begin{equation}
\label{eqn:draw2}
D_{\rm raw} = (1-F^{\: 0}_{{\rm fake} \ \ell}) \frac{ (1 - 2 \bar{\chi}')(1 - 2 f^{\rm ws}_{\rm seq}) 
          (N^{\bbbar}_T - N^{\bbbar}_M)  +  (N^{\ccbar}_T - N^{\ccbar}_M) }
   {  N^{\bbbar}_T + N^{\bbbar}_M   +   N^{\ccbar}_T + N^{\ccbar}_M  }
\end{equation}

Using Equation~\ref{eqn:appd}, we define the true flavor tagging dilution 
in $\bbbar$ and $\ccbar$ events as
\begin{equation}
\label{eqn:dbbdef}
D_{\bbbar} \equiv \frac{N^{\bbbar}_T - N^{\bbbar}_M}{N^{\bbbar}_T + N^{\bbbar}_M}
\end{equation}
and
\begin{equation}
\label{eqn:dccdef}
D_{\ccbar} \equiv \frac{N^{\ccbar}_T - N^{\ccbar}_M}{N^{\ccbar}_T + N^{\ccbar}_M}
\end{equation}
respectively.  The fraction of events from $\bbbar$ and $\ccbar$ production
are defined by
\begin{equation}
\label{eqn:fbbdef}
 F_{\bbbar} \equiv \frac{ N^{\bbbar}_T + N^{\bbbar}_M }
                     {  N^{\bbbar}_T + N^{\bbbar}_M   +   N^{\ccbar}_T + N^{\ccbar}_M }
\end{equation}
and 
\begin{equation}
\label{eqn:fccdef}
 F_{\ccbar} \equiv \frac{ N^{\ccbar}_T + N^{\ccbar}_M }
                     {  N^{\bbbar}_T + N^{\bbbar}_M   +   N^{\ccbar}_T + N^{\ccbar}_M }
\end{equation}
respectively.
Combining Equations~\ref{eqn:draw1}, \ref{eqn:dbbdef}, \ref{eqn:dccdef}, 
\ref{eqn:fbbdef}, \ref{eqn:fccdef} gives
\begin{eqnarray}
D_{\rm raw} & = & (1 - F^{\: 0}_{{\rm fake} \ \ell} ) \left[
       (1 - 2 \bar{\chi}')(1 - 2 f^{\rm ws}_{\rm seq}) \ F_{\bbbar} \ D_{\bbbar}
                 + F_{\ccbar} \ D_{\ccbar}  \right] \nonumber \\
D_{\rm raw} & = & \left\{ (1 - F^{\: 0}_{{\rm fake} \ \ell} ) \left[
       (1 - 2 \bar{\chi}')(1 - 2 f^{\rm ws}_{\rm seq}) \ F_{\bbbar} 
            + F_{\ccbar} \ \frac{D_{\ccbar}}{D_{\bbbar}}  \right] \right\} D_{\bbbar} \nonumber\\ 
D_{\rm raw} & = & \frac{1}{N_D} D_{\bbbar}
\label{eqn:draw3}
\end{eqnarray}
where we have defined the dilution normalization factor $N_D$ as
\begin{equation}
\label{eqn:nddef}
  \frac{1}{N_D} \equiv (1 - F^{\: 0}_{{\rm fake} \ \ell} ) \left[
     (1 - 2 \bar{\chi}')(1 - 2 f^{\rm ws}_{\rm seq}) \ F_{\bbbar} 
            + F_{\ccbar} \ \frac{D_{\ccbar}}{D_{\bbbar}}  \right].
\end{equation}
Equation~\ref{eqn:nddef} can be used to calculate the expected values
for the $N_D$ parameters.  For this calculation, we will assume
\begin{itemize}
  \item $\bar{\chi}' \approx 0.20$ from the Monte Carlo. 
  \item $f^{\rm ws}_{\rm seq}(e{\rm -trigger}) = 0.07$ and
        $f^{\rm ws}_{\rm seq}(\mu{\rm -trigger}) = 0.10$ using
        $f^{\rm ws}_{\rm seq} = 0.75 \times f_{\rm seq}$ and the
        values in Table~\ref{tab:fitparam}.
  \item The $F_{\bbbar}$ values are given in Table~\ref{tab:mclptrel}.
        We also use $F_{\ccbar} = 1 - F_{\bbbar}$.
  \item For the JCT, we assume $D_{\ccbar}/D_{\bbbar}({\rm JCT}) = 0.5$.
        Using the average SLT dilution and the assumption
        that $D_{\ccbar}({\rm SLT}) = 0.5$, we estimate
        $D_{\ccbar}/D_{\bbbar}({\rm SLT}) \approx 1.3$.
\end{itemize}
Using the numbers above, we find
\begin{itemize}
  \item $N^e_{D,{\rm SLT}} \approx 1.8$.
  \item $N^\mu_{D,{\rm SLT}} \approx 2.1$.
  \item $N^e_{D,{\rm JCSV}} \approx N^e_{D,{\rm JCDV}} \approx 1.9$.
  \item $N^\mu_{D,{\rm JCSV}} \approx N^\mu_{D,{\rm JCDV}} \approx 2.4$.
\end{itemize}

\newpage


\section{ Fraction of Fake Trigger Muons }
\label{app:fake_muons}

As is stated in Section~\ref{sec:fakeMu}, we believe that 
most of the fake trigger-muons in the data are from
heavy flavor decay.
There may be some correlation on average between the sign fake muon 
charge and the $B$ flavor at decay, however we assume that
this correlation is smaller than that of real trigger-muons.
Fake muon events are divided into two groups:
\begin{enumerate}
  \item Fake muon events whose dilution is the same as the
        dilution for real muons.
  \item Fake muon events whose dilution is zero.
\end{enumerate}
We treat group~1 as if they are real trigger-muon events.
We treat group~2 as if they are $b\bar{b}$ events with a 
flavor tagging dilution of zero.

We determine the fraction of events with fake muons that have zero
dilution in the $\mu$-trigger data by assuming 
that the {\em true} flavor tagging dilution is
the same in the $e$ and $\mu$ trigger data.
The raw dilution ($D_{\rm raw}$), which assumes all opposite-sign (same-sign)
events are tags (mistags), is different for the
$e$ and $\mu$ trigger data for the following reasons:
\begin{enumerate}
  \item The fraction of real trigger-leptons in $b\bar{b}$
        events that are not from direct $b\rightarrow \ell$ decay
        ($f_{\rm seq}$) is 9.4\% for the $e$-trigger data
        and 13.6\% for the $\mu$-trigger data.
  \item The fraction of events from $\ccbar$ production $F_{\ccbar}$ 
        is slightly different (see Table~\ref{tab:mclptrel}).
  \item We estimate that only 1\% of the trigger electrons
        are fake, while, as shown below, about 10\% of the $\mu$-trigger events
        contain a fake muon.
\end{enumerate}
We can correct $D_{\rm raw}$ for (1) and (2) using the equation
\begin{equation}
    \langle D_{\rm raw}' \rangle = \frac{ \langle D_{\rm raw} \rangle }
      { F_{b\bar{b}} \cdot (1 - 2 f_{\rm seq}^{\rm ws} + 
        F_{c\bar{c}} \cdot D_{c\bar{c}}/D_{b\bar{b}})  }
\end{equation}
where $f_{\rm seq}^{\rm ws}$ is the fraction of non $b\rightarrow \ell$
decays that have the ``wrong'' sign.  The Monte Carlo gives
$f_{\rm seq}^{\rm ws} = 0.75 \cdot f_{\rm seq}$.
The values of $\langle D'_{\rm raw}\rangle$ for the 
SLT, JCSV, and JCDV flavor tagging methods
in the $e$ and $\mu$-trigger data are given
in Table~\ref{tab:dprime}.
The weighted average of 
$\langle D'_{\rm raw}(e)\rangle/\langle D'_{\rm raw}(\mu)\rangle$
for the SLT, JCSV, and JCDV flavor tagging methods
gives 1.14 $\pm$ 0.08. 
The fraction of events with fake muons that have zero dilution
can be extracted using
\begin{equation}
\label{eqn:FfakeMu}
  F^{\: 0}_{{\rm fake} \: \mu} = 1 - \frac{1}
     {\langle D_{\rm raw}'(e)\rangle / \langle D_{\rm raw}'(\mu) \rangle}.
\end{equation}
Equation~\ref{eqn:FfakeMu} gives $F^{\: 0}_{{\rm fake} \: \mu} = 12 \pm 6$ \%. 
The relatively large uncertainty on $F^{\: 0}_{{\rm fake} \: \mu}$ 
gives a significant systematic uncertainty on the dilution normalization
$N_D$ for the flavor tags (see Table~\ref{tab:ndsys_m}),
however, the contribution to the systematic uncertainty on $\Delta m_d$ 
is relatively small.

\begin{table}
\caption{
The fraction of events from $b\bar{b}$ production $F_{\bbbar}$
for the $e$ and $\mu$ trigger data.
The first two columns give the result for $F_{\bbbar}$
from fitting the $\ptrel$ and $\mcl$ spectra to a
linear combination of $\bbbar$ and $\ccbar$ templates.
The error in the first two columns is statistical only.
The last column gives the average of the $\ptrel$ and $\mcl$
results.
The error on the average takes into account the difference
in the $\ptrel$ and $\mcl$ fit results  as a systematic error.
The fraction of events from 
$\ccbar$ production is given by $F_{\ccbar}=1-F_{\bbbar}$.
}
\label{tab:mclptrel}
\begin{tabular}{lccc}
Sample  &  $\ptrel$ Fit  &  $\mcl$ Fit  &  Average Value  \\
\tableline
 JCSV($e$)    &  89.6 $\pm$ 0.6 \%  &  92.2 $\pm$ 0.5 \%  &  90.9 $\pm$ 1.3 \%  \\
 JCDV($e$)    &  94.5 $\pm$ 1.5 \%  &  96.7 $\pm$ 1.0 \%  &  95.6 $\pm$ 1.5 \%  \\
 SLT($e$)     &  91.3 $\pm$ 1.7 \%  &  94.4 $\pm$ 1.5 \%  &  92.9 $\pm$ 1.7 \%  \\
\tableline
 JCSV($\mu$)  &  90.1 $\pm$ 0.5 \%  &  89.0 $\pm$ 0.5 \%  &  89.6 $\pm$ 1.3 \%  \\
 JCDV($\mu$)  &  98.7 $\pm$ 1.3 \%  &  97.6 $\pm$ 1.2 \%  &  98.2 $\pm$ 1.3 \%  \\
 SLT($\mu$)   &  92.8 $\pm$ 1.6 \%  &  91.5 $\pm$ 1.4 \%  &  92.2 $\pm$ 1.6 \%  \\ 
\end{tabular}
\end{table}
\begin{table}
\caption{
   Fitted values for the SLT raw dilution parameterization constants
   ($A$ and $B$) and the raw dilution for isolated soft leptons
   (no-$\ptrel$) for soft $e$ and $\mu$ tags in the 
   $e$ and $\mu$ trigger data.
   The errors are statistical.
   The SLT raw dilution is parameterized as a function of the
   soft lepton $\ptrel$ with the functional form
   $D_{\rm raw}(\ptrel) = A \cdot \left( 1 - e^{ - \ptrel + B } \right)$
   where $\ptrel$ is in GeV/$c$. 
}
\label{tab:sltAB}
\begin{tabular}{ccccc}
  & \multicolumn{2}{c}{ $e$-Trigger}  
  & \multicolumn{2}{c}{ $\mu$-Trigger}  \\
  & Soft-$e$  &  Soft-$\mu$ 
  & Soft-$e$  &  Soft-$\mu$  \\
\tableline
  $A$  &  0.41 $\pm$ 0.05  &  0.46 $\pm$ 0.03  &  0.42 $\pm$ 0.05  &  0.36 $\pm$ 0.03  \\
  $B$  &  0.31 $\pm$ 0.13  &  0.43 $\pm$ 0.06  &  0.35 $\pm$ 0.11  &  0.25 $\pm$ 0.09  \\
no-$\ptrel$  &  0.36 $\pm$ 0.04  &  0.25 $\pm$ 0.03  &  0.18 $\pm$ 0.04  &  0.18 $\pm$ 0.03 \\
\end{tabular}
\end{table}

\begin{table}
\caption{ 
   Constants for the jet-charge tag raw dilution parameterization
   as a function of $|Q_{\rm jet}|$.
   The parameterization has the form $D_{\rm raw} = D_{\rm max} \cdot |Q_{\rm jet}|$
   except for events where $|Q_{\rm jet}| = 1$, in which case
   the average $D_{\rm raw}$ is used.
 }
\label{tab:jcDmax}
\begin{tabular}{lcccc}
     & \multicolumn{2}{c}{ $e$-Trigger}  
     & \multicolumn{2}{c}{ $\mu$-Trigger}  \\  
     &  JCSV    &    JCDV     &    JCSV      &   JCDV   \\
\tableline
  $D_{\rm max}$   &  0.083 $\pm$ 0.010  
                  &  0.34  $\pm$ 0.02  
                  &  0.060 $\pm$ 0.009
                  &  0.29  $\pm$ 0.02  \\
  $D_{\rm raw}$, $|Q_{\rm jet}|=1$  
                  &  0.091 $\pm$ 0.014
                  &  0.18  $\pm$ 0.03 
                  &  0.074 $\pm$ 0.014
                  &  0.12  $\pm$ 0.03 \\
\end{tabular}
\end{table}

\begin{table}
{\caption
{\label{tab:fitparam}
The fixed input parameters used in the fit.
The first column lists the parameters, which are described below.
The second column lists the value of the parameter assumed for the fit,
with an error on this parameter, if appropriate.
The third column lists the variation of the parameter for the purposes
of evaluating the systematic errors on $\Delta m_d$ and $N_D$;
$\pm 1\sigma$ means that the parameter was varied by the error listed
in the second column.
Finally the fourth column lists the source of the parameter.
The fixed input parameters are
the $B$ hadron lifetimes ($\tau_i$),
the $B^0_s$ mass difference ($\Delta m_s$),
the relative $B$ hadron production
fractions ($f_i$),
the fraction of trigger leptons from $b\bar{b}$ sources other
than direct $B$ decay ($f_{seq}$),
the flavor tag dilution value for $c\bar{c}$ events ($D_{c\bar{c}}$),
an effective lifetime for $c\bar{c}$ events ($\tau_{c\bar{c}}$),
and the fraction of $\mu$-trigger events where the trigger 
muon is really a hadron and has a dilution of zero ($F^{\: 0}_{{\rm fake} \: \mu}$).
}}
\begin{tabular}{llll}
   Parameter       &   Value  & Variation &  Source \\
\tableline
\tableline
$\tau_{B^0}$                   &
$1.56\pm 0.06$~ps              & $\pm 1 \sigma$ & \cite{PDG} \\
$\tau_{B^+}/\tau_{B^0}$        &
$1.02\pm 0.05$                 & $\pm 1 \sigma$ & \cite{PDG} \\
$\tau_{B^0_s}$                &
$1.61^{+0.10}_{-0.09}$~ps      & $\pm 1 \sigma$ & \cite{PDG} \\
$\tau_{\Lambda_b}$            & 
$1.14\pm 0.08$~ps              & $\pm 1 \sigma$ & \cite{PDG} \\
\tableline
$\Delta m_s$                   &
700 $\hbar{\rm ps}^{-1}$      & down to 6.7 $\hbar{\rm ps}^{-1}$ & Assumption\\
\tableline
$f_{B^+}$                      &
$37.8\pm 2.2$~\%               & $\pm 1 \sigma$ & \cite{PDG} \\
$f_{B^0}$                      &
$37.8\pm2.2$~\%                & $\pm 1 \sigma$ & \cite{PDG} \\
$f_{B^0_s}$                    &
$11.1^{+2.5}_{-2.6}$~\%        & $\pm 1 \sigma$ & \cite{PDG} \\
$f_{\Lambda_b}$                &
$13.2\pm4.1$~\%                & $\pm 1 \sigma$ & \cite{PDG} \\
\tableline
$f_{\rm seq}$                  &
$e$:   \ \  9.4  \%            & $\pm 0.25 \times f_{\rm seq}$ & Monte Carlo\\
                               &
$\mu$:  13.6  \%               & $\pm 0.25 \times f_{\rm seq}$ & Monte Carlo\\
\tableline
JCT $D_{c\bar{c}}/D_{b\bar{b}}$ &
0.5                            & 0 to 1         & Assumption \\
SLT $D_{c\bar{c}}$             &
0.5                            & 0 to 1         & Assumption \\
$\tau_{c\bar{c}}$              &
$1.53\pm 0.20$~ps              & $\pm 1 \sigma$ & fit to $c\bar{c}$ MC  \\
\tableline
$F^{\: 0}_{{\rm fake} \: \mu}$        &
$12\pm 6$~\%            &  $\pm 1 \sigma$ & $e/\mu$ $D_{\rm raw}$ comparison \\
\end{tabular}
\end{table}

\begin{table}
\caption{ 
   Results of the maximum likelihood fit for $\Delta m_d$ and 
   the flavor tag dilution normalization factors for the
   individual flavor taggers separately in the $e$ and $\mu$ trigger data.
   The first error is statistical and the second is systematic.
   The evaluation of the systematic errors is discussed later
   in the text.
   The dilution normalization factors are given for two 
   cases: $\Delta m_d$ free to float in the fit and
   $\Delta m_d$ fixed to the `96 world average [18]
   ( 0.474 $\hbar \ {\rm ps}^{-1}$).  
 }
\label{tab:indivFits}
\begin{tabular}{lccc}
     &  SLT  &  JCSV  &  JCDV  \\
\tableline
  $e$-trigger  & & & \\
 $N_D$, $\Delta m_d$ free  &  1.66 $\pm$ 0.12 $\pm$ 0.18  
                           &  1.84 $\pm$ 0.23 $\pm$ 0.19
                           &  1.71 $\pm$ 0.16 $\pm$ 0.11 \\
 $N_D$, $\Delta m_d$ fixed &  1.72 $\pm$ 0.08 $\pm$ 0.11
                           &  1.88 $\pm$ 0.20 $\pm$ 0.15
                           &  1.76 $\pm$ 0.13 $\pm$ 0.09 \\
 $\Delta m_d$ ($\hbar \ {\rm ps}^{-1}$)  &  0.45 $\pm$ 0.08 $\pm$ 0.05 
                     & \multicolumn{2}{c}{  0.42 $\pm$ 0.09 $\pm$ 0.03  } \\
\tableline
  $\mu$-trigger  & & & \\
 $N_D$, $\Delta m_d$ free  &  2.05 $\pm$ 0.19 $\pm$ 0.37  
                           &  2.86 $\pm$ 0.40 $\pm$ 0.43
                           &  2.52 $\pm$ 0.28 $\pm$ 0.25 \\
 $N_D$, $\Delta m_d$ fixed &  2.01 $\pm$ 0.13 $\pm$ 0.22
                           &  2.41 $\pm$ 0.29 $\pm$ 0.39 
                           &  2.14 $\pm$ 0.33 $\pm$ 0.25 \\
 $\Delta m_d$ ($\hbar \ {\rm ps}^{-1}$)  &  0.50 $\pm$ 0.09 $\pm$ 0.05 
                     & \multicolumn{2}{c}{  0.68 $\pm$ 0.11 $\pm$ 0.04  } \\
\end{tabular}
\end{table}
\begin{table}
\caption{
Results of the maximum likelihood fit for $\Delta m_d$
and the flavor tag dilution normalization parameters ($N_D$)
for the $e$ and $\mu$ trigger data using both soft-lepton
and jet-charge flavor tagging.
The first error is the statistical error on the parameter,
determined by the fit,
and the second error is the systematic error,
which is determined using the prescription described in the text.
  The dilution normalization parameters ($N_D$) for the JCT
  are lower than those in Table~\ref{tab:indivFits}
  because the SLT is used for double flavor tagged events (SLT and JCT),
  which lowers the average dilution for the JCT.
Only the statistical errors are given for the dilution normalization
parameters.
}
\label{tab:fit_results}
\begin{tabular}{lc}
Parameter                         & Fit Result                          \\ \hline
$\Delta m_d$ ($\hbar \ {\rm ps}^{-1}$)  & $0.500 \pm 0.052 \pm 0.043$         \\ \hline
$N_{D,{\rm JCSV}}^e$                  & $1.43 \pm 0.22 $           \\
$N_{D,{\rm JCDV}}^e$                  & $1.53 \pm 0.16 $           \\
$N_{D,{\rm SLT}}^e$                   & $1.72 \pm 0.10 $           \\ \hline
$N_{D,{\rm JCSV}}^\mu$                & $1.99 \pm 0.32 $           \\
$N_{D,{\rm JCDV}}^\mu$                & $2.00 \pm 0.22 $           \\
$N_{D,{\rm SLT}}^\mu$                 & $2.05 \pm 0.16 $           \\
\end{tabular}
\end{table}
\begin{table}
\caption{ 
    The individual contributions to the systematic error on $\Delta m_d$.
    The value of $\delta \Delta m_d$ is the amount by which the fitted
    value of $\Delta m_d$ shifted when a parameter was varied.
    The size of the variations are given in Table~\ref{tab:fitparam}
    and in the text.
    The parameters that were varied are: 
      the $B$ hadron lifetimes ($\tau_i$),
      the assumed value of $\Delta m_s$,
      the $B^0_s$ and $\Lambda_b$ production fractions ($f_i$),
      the fraction of trigger-leptons from sequential decay ($f_{\rm seq}$),
      the assumed dilution for $\ccbar$ events,
      the effective lifetime for $\ccbar$ events ($\tau_{\ccbar}$),
      the fraction of trigger muons that are fake and
            have zero dilution ($F^{\: 0}_{{\rm fake} \ \mu}$),
      the normalization for the fake-vertex component,
      the assumed dilution for fake-vertex events ($D_{\rm FV}/D_{\bbbar}$),
      the fraction of $\ccbar$ events $F_{\ccbar}$,
      and the parameterization of $D_{\rm raw}$.
    The total systematic error is the sum of the individual contributions
    in quadrature.
}
\label{tab:dmdsys}
\begin{tabular}{lc}
Parameter &  \multicolumn{1}{c}{ $\delta \Delta m_d \ (\hbar \ {\rm ps}^{-1})$ } \\
\tableline
   $\tau_{B^0}$                  &  0.004  \\
   $\tau_{B^+}/\tau_{B^0}$        &  0.021  \\
   $\tau_{B^0_s}$                &  0.005  \\
   $\tau_{\Lambda_b}$            &  0.005  \\
\tableline
   $\Delta m_s$                  &  0.004  \\
   $f_{B^0_s}$                   &  0.001  \\
   $f_{\Lambda_b}$               &  0.007  \\
\tableline
   $f_{\rm seq}$                 &  0.004  \\
   JCT $D_{\ccbar}/D_{\bbbar}$   &  0.002  \\
   SLT $D_{\ccbar}$              &  0.032  \\
   $\tau_{\ccbar}$               &  0.006  \\
\tableline
   $F^{\: 0}_{{\rm fake} \ \mu}$        &  0.001  \\
   Fake-vertex component         &  0.006  \\
   $D_{\rm FV}/D_{\bbbar}$       &  0.003  \\
\tableline
   $F_{\ccbar}$, $e$-trigger     &  0.006  \\
   $F_{\ccbar}$, $\mu$-trigger   &  0.003  \\
\tableline
   JCT $D_{\rm raw}$ parameterization      &  0.003  \\
   SLT $D_{\rm raw}$ parameterization      &  0.009  \\
\tableline
   Total systematic error        &  0.043   \\
\end{tabular}
\end{table}
\begin{table}
\caption{ 
    The individual contributions to the systematic error on the
    dilution normalization parameters for the
    $e$-trigger data $N^e_{D,i}$.
    The value of $\delta N^e_{D,i}$ is the amount by which the fitted
    value of $N^e_{D,i}$ shifted when a parameter was varied.
    The size of the variations are given in Table~\ref{tab:fitparam}
    and in the text.
    These systematic errors are for the case where $\Delta m_d$ is
    allowed to float in the fit.
    The parameters that were varied are: 
      the $B$ hadron lifetimes ($\tau_i$),
      the assumed value of $\Delta m_s$,
      the $B^0_s$ and $\Lambda_b$ production fractions ($f_i$),
      the fraction of trigger-leptons from sequential decay ($f_{\rm seq}$),
      the assumed dilution for $\ccbar$ events,
      the effective lifetime for $\ccbar$ events ($\tau_{\ccbar}$),
      the normalization for the fake-vertex component,
      the assumed dilution for fake-vertex events ($D_{\rm FV}/D_{\bbbar}$),
      the fraction of $\ccbar$ events $F_{\ccbar}$,
      and the parameterization of $D_{\rm raw}$.
    The total systematic error is the sum of the individual contributions
    in quadrature.
        }
\label{tab:ndsys_e}
\begin{tabular}{lccc}
Parameter   &  \multicolumn{1}{c}{ $\delta N^e_{D, {\rm JCSV}}$ } 
            &  \multicolumn{1}{c}{ $\delta N^e_{D,{\rm JCDV}}$ }
            &  \multicolumn{1}{c}{ $\delta N^e_{D,{\rm SLT}}$  }\\
\tableline
  $\tau_{B^0}$                 &  0.01  &  0.01  &  0.01  \\
  $\tau_{B^+}/\tau{B^0}$       &  0.01  &  0.01  &  0.01  \\
  $\tau_{B^0_s}$               &  0.01  &  0.01  &  0.01  \\
  $\tau_{\Lambda_b}$           &  0.00  &  0.00  &  0.00  \\
\tableline
  $\Delta m_s$                 &  0.02  &  0.01  &  0.01  \\
  $f_{B^0_s}$                  &  0.05  &  0.05 &  0.05 \\
  $f_{\Lambda_b}$              &  0.01  &  0.01  &  0.01  \\
\tableline
  $f_{\rm seq}$                &  0.06 &  0.06 &  0.06 \\
   Assumed $\ccbar$ dilution   &  0.12  &  0.05  &  0.15  \\
   $\tau_{\ccbar}$             &  0.00  &  0.00  &  0.01  \\
\tableline
   Fake-vertex component       &  0.02  &  0.01  &  0.02  \\
   $D_{\rm FV}/D_{\bbbar}$     &  0.06  &  0.03  &  0.04  \\
\tableline
   $F_{\ccbar}$                &  0.00  &  0.01  &  0.01  \\
\tableline
   $D_{\rm raw}$ parameterization     &  0.10  &  0.05  &  0.06  \\
\tableline
\multicolumn{1}{l}{ Total Systematic Error }      
     &   0.19   &   0.11   &   0.18   \\
\end{tabular}
\end{table}
\begin{table}
\caption{ 
    The individual contributions to the systematic error on the
    dilution normalization parameters for the
    $\mu$-trigger data $N^\mu_{D,i}$.
    The value of $\delta N^\mu_{D,i}$ is the amount by which the fitted
    value of $N^\mu_{D,i}$ shifted when a parameter was varied.
    The size of the variations are given in Table~\ref{tab:fitparam}
    and in the text.
    These systematic errors are for the case where $\Delta m_d$ is
    allowed to float in the fit.
    The parameters that were varied are: 
      the $B$ hadron lifetimes ($\tau_i$),
      the assumed value of $\Delta m_s$,
      the $B^0_s$ and $\Lambda_b$ production fractions ($f_i$),
      the fraction of trigger-leptons from sequential decay ($f_{\rm seq}$),
      the assumed dilution for $\ccbar$ events,
      the effective lifetime for $\ccbar$ events ($\tau_{\ccbar}$),
      the fraction of trigger muons that are fake and have
             a dilution of zero ($F^{\: 0}_{{\rm fake} \ \mu}$),
      the normalization for the fake-vertex component,
      the assumed dilution for fake-vertex events ($D_{\rm FV}/D_{\bbbar}$),
      the fraction of $\ccbar$ events $F_{\ccbar}$,
      and the parameterization of $D_{\rm raw}$.
    The total systematic error is the sum of the individual contributions
    in quadrature.
        }
\label{tab:ndsys_m}
\begin{tabular}{lccc}
Parameter   &  \multicolumn{1}{c}{ $\delta N^\mu_{D, {\rm JCSV}}$ } 
            &  \multicolumn{1}{c}{ $\delta N^\mu_{D,{\rm JCDV}}$ }
            &  \multicolumn{1}{c}{ $\delta N^\mu_{D,{\rm SLT}}$  }\\
\tableline
  $\tau_{B^0}$                      &  0.00  &  0.00  &  0.01  \\
  $\tau_{B^+}/\tau{B^0}$            &  0.02  &  0.02  &  0.00  \\
  $\tau_{B^0_s}$                    &  0.01  &  0.01  &  0.01  \\
  $\tau_{\Lambda_b}$                &  0.02  &  0.02  &  0.01  \\
\tableline
  $\Delta m_s$                      &  0.02  &  0.01  &  0.01  \\
  $f_{B^0_s}$                       &  0.08  &  0.07  &  0.06  \\
  $f_{\Lambda_b}$                   &  0.02  &  0.01  &  0.01  \\
\tableline
  $f_{\rm seq}$                     &  0.15  &  0.14  &  0.13  \\
  Assumed $\ccbar$ dilution         &  0.28  &  0.03  &  0.26  \\
   $\tau_{\ccbar}$                  &  0.01  &  0.00  &  0.01  \\
\tableline
   $F^{\: 0}_{{\rm fake} \ \mu}$           &  0.19  &  0.17  &  0.15  \\
   Fake vertex component            &  0.00  &  0.01  &  0.01  \\
   $D_{\rm FV}/D_{\bbbar}$          &  0.13  &  0.08  &  0.06  \\
\tableline
   $F_{\ccbar}$                     &  0.00  &  0.01  &  0.02  \\
\tableline
   $D_{\rm raw}$ parameterization   &  0.17  &  0.03  &  0.07  \\
\tableline
\multicolumn{1}{l}{ Total Systematic Error }      
     &   0.43  &  0.25  &  0.37  \\
\end{tabular}
\end{table}
\begin{table}
\caption{ The statistical power $\epsilon D^2$ 
          for the flavor tagging methods used:
          Jet-Charge Single Vertex (JCSV),
          Jet-Charge Double Vertex (JCDV),
          and Soft-Lepton Tag (SLT).
          Results for the $e$ and $\mu$ trigger
          data are shown in separate rows.
          The sum is over bins of $\ptrel$ 
          for the soft-lepton data and $|Q_{\rm jet}|$ for
          the jet-charge data, as shown in Figures~\ref{fig:sltDraw}
          and~\ref{fig:dplot}, respectively.  
          The square of the dilution normalization factor $N_D$ is used 
          to rescale the $\sum_i \epsilon_i D_{{\rm raw} \ i}^2$
          value to give $\sum_i \epsilon_i D_i^2$.
          The first error is statistical, the second systematic.
          }
\label{tab:eds}
\begin{tabular}{lrccc}
Sample  &  \multicolumn{1}{c}{ Total $\epsilon$ }  
        &  $\sum_i \epsilon_i D_{{\rm raw} \ i}^2$  
        &  $N_D$
        &  $\sum_i \epsilon_i D_i^2$  \\
\tableline
 JCSV ($e$)    &  41.55  $\pm$ 0.14  \%  
                           &   0.077 $\pm$ 0.016 \%  
                           &   1.88  $\pm$ 0.20 $\pm$ 0.15
                           &   0.27  $\pm$ 0.06 $\pm$ 0.04 \% \\
 JCDV ($e$)    &   7.44  $\pm$ 0.08  \%  
                           &   0.159 $\pm$ 0.023 \%  
                           &   1.76  $\pm$ 0.13 $\pm$ 0.09 
                           &   0.49  $\pm$ 0.10 $\pm$ 0.05 \% \\
 SLT      ($e$)    &   4.38  $\pm$ 0.06  \%  
                           &   0.329 $\pm$ 0.033 \% 
                           &   1.72  $\pm$ 0.08 $\pm$ 0.11
                           &   0.97  $\pm$ 0.13 $\pm$ 0.12 \% \\
\tableline
 JCSV ($\mu$)  &  43.81  $\pm$ 0.14  \%  
                           &   0.048 $\pm$ 0.012 \%  
                           &   2.41  $\pm$ 0.29 $\pm$ 0.39 
                           &   0.28  $\pm$ 0.06 $\pm$ 0.05 \% \\
 JCDV ($\mu$)  &   7.66  $\pm$ 0.07  \%  
                           &   0.113 $\pm$ 0.018 \%  
                           &   2.14  $\pm$ 0.33 $\pm$ 0.25 
                           &   0.52  $\pm$ 0.18 $\pm$ 0.12 \% \\
 SLT      ($\mu$)  &   4.54  $\pm$ 0.06  \%  
                           &   0.210 $\pm$ 0.026 \%  
                           &   2.01  $\pm$ 0.13 $\pm$ 0.22 
                           &   0.85  $\pm$ 0.15 $\pm$ 0.19 \% \\
\end{tabular}
\end{table}

\begin{table}
\caption{ Average raw dilution for three flavor taggers in the
electron and muon data corrected for wrong sign sequential
decay trigger leptons and the $b\bar{b}$ to $c\bar{c}$ ratio.  }
\label{tab:dprime}
\begin{tabular}{lrrrr}
  \multicolumn{1}{c}{ Flavor Tag }  &
  \multicolumn{1}{c}{ $\langle D_{\rm raw} \rangle$ } &
  \multicolumn{1}{c}{ Correction }   &
  \multicolumn{1}{c}{ $\langle D_{\rm raw}'\rangle$ } &
  \multicolumn{1}{c}{ $ \langle D_{\rm raw}'(e)\rangle /
         \langle D_{\rm raw}'(\mu)\rangle   $  } \\
\tableline \tableline
JCSV ($e$)    &  2.7 $\pm$ 0.5 \%  &  1.16  &  3.1 $\pm$ 0.6 \%  &
    1.07 $\pm$ 0.28  \\
JCSV ($\mu$)  &  2.3 $\pm$ 0.4 \%  &  1.26  &  2.9 $\pm$ 0.5 \%  &
                     \\
\tableline
JCDV ($e$)    & 12.1 $\pm$ 1.1 \%  &  1.15  & 14.2 $\pm$ 1.3 \%  &
    1.18 $\pm$ 0.16  \\
JCDV ($\mu$)  &  9.7 $\pm$ 1.0 \%  &  1.24  & 12.0 $\pm$ 1.2 \%  &
                     \\
\tableline
SLT ($e$)       & 22.1 $\pm$ 1.3 \%  &  1.15  & 25.5 $\pm$ 1.5 \%  &
    1.13 $\pm$ 0.10  \\
SLT ($\mu$)     & 18.2 $\pm$ 1.3 \%  &  1.24  & 22.5 $\pm$ 1.6 \%  &
                     \\
\end{tabular}
\end{table}


\begin{thebibliography}{99}

  \bibitem{SM} S.~L.~Glashow, Nucl.~Phys.~{\bf 22}, 579 (1961):\\
    S.~Weinberg, Phys.~Rev.~Lett.~{\rm 19}, 1264 (1967);\\
    A.~Salam, {\em Elementary Particle Theory}, ed.~N.~Svartholm,
    Almquist and Wiksells, Stockholm (1969) p.~367.

  \bibitem{ckm} N. Cabibbo, Phys. Rev. Lett., {\bf 10}, 531 (1963). \\
                M. Kobayashi and K. Maskawa, Prog. Theor. Phys. {\bf 49}, 652 (1973).


  \bibitem{TDR} CDF Collaboration, R. Blair {\it et al}. The CDFII Detector 
      Technical Design Report, FERMILAB-PUB-96/390-E.

  \bibitem{ref:cdf_b_refs}

	CDF Collaboration, F.~Abe {\it et al.},
	Phys.~Rev.~D {\bf 59}, 032004 (1999).\\
	CDF Collaboration, F.~Abe {\it et al.},
	Phys.~Rev.~D {\bf 58}, 112004 (1998).\\
	CDF Collaboration, F.~Abe {\it et al.},
	Phys.~Rev.~D {\bf 58}, 092002 (1998).\\
	CDF Collaboration, F.~Abe {\it et al.},
	Phys.~Rev.~D {\bf 57}, 5382 (1998).\\
	CDF Collaboration, F.~Abe {\it et al.},
	Phys.~Rev.~D {\bf 55}, 1142 (1997).\\
	CDF Collaboration, F.~Abe {\it et al.},
	Phys.~Rev.~Lett.~{\bf 77}, 1945 (1997).\\
	CDF Collaboration, F.~Abe {\it et al.},
	Phys.~Rev.~D {\bf 53}, 3496 (1996).

  \bibitem{ribon} CDF Collaboration, F.~Abe {\it et al.},
{\it Measurement of the $B^0_d\bar{B}^0_d$ Oscillation Frequency Using
Dimuon Data in $p\bar{p}$ Collisions at $\sqrt{s}=1.8$~TeV},
FERMILAB-Pub-99/030-E, submitted to Physical Review Letters.


  \bibitem{petar}

        CDF Collaboration, F.~Abe {\it et al}.,
        Phys.~Rev.~Lett., {\bf 80}, 2057, (1998);\\
	CDF Collaboration, F.~Abe {\it et al}.,
        Phys.~Rev.~D {\bf 59}, 032001 (1999).


  \bibitem{PDG98} C. Caso {\it et al.}  (The Particle Data Group),
        The European Physics Journal {\bf C3} (1998) 1.


  \bibitem{sin2beta_prl}

	CDF Collaboration, F.~Abe {\it et al}.,
        Phys.~Rev.~Lett., {\bf 81}, 5513, (1998).

  \bibitem{MixResults}  
	A summary of published measurements and their average can be found in 
        C.~Caso et al, The European Physical Journal {\bf C3}, 1 (1998).


  \bibitem{JetCharge}  
        ALEPH Collaboration, D.~Buskulic {\it et al.},
        Zeit.~Phys.~C {\bf 75}, 397 (1997);\\
        DELPHI Collaboration, P.~Abreu {\it et al.},
        Zeit.~Phys.~C {\bf 72}, 17 (1996);\\
        OPAL Collaboration, R.~Akers {\it et al.},
        Phys.~Lett.~B {\bf 327}, 411 (1994);\\
        SLD Collaboration, K.~Abe {\it et al.},
        Phys.~Rev.~Lett.~{\bf 74}, 2890 (1995).

  \bibitem{detector} F. Abe {\it et al.}, Nucl. Instrum. Methods
        Phys. Res. A {\bf 271}, 387 (1988).

  \bibitem{TopPRD} 
     F.~Abe {\it et al.}, Phys.~Rev.~Lett.~{\bf 74} (1995) 2626;\\
     see also F.~Abe {\it et al.}, Phys.~Rev.~D~{\bf 50} (1994) 2966.

  \bibitem{SVX} D.~Amidei {\it et al.},  Nucl. Instr. Methods 
      {\bf A350}, 73, (1994); \\
    P.~Azzi {\it et al.}, Nucl. Instr. Methods {\bf A360}, 137 (1995).

  \bibitem{cft}
     G.~W.~Foster {\it et al.}, Nucl.~Instrum.~Methods~A {\bf 269},
     93 (1988).


  \bibitem{deltaR}
  $\Delta R = \sqrt{ (\Delta \phi)^2 + (\Delta \eta)^2 }$
  where $\phi$ is the azimuthal 
  direction with respect
  to the beam and $\eta$ is the pseudorapidity,
  which is defined as $\eta = - \ln ( \tan (\frac{\theta}{2}))$. 

  \bibitem{Pythia} H. U. Bengtsson and T. Sj\"{o}strand, Computer
        Physics Commun. {\bf 43}, 43 (1987); T.~Sj\"{o}strand, Computer
        Physics Commun. {\bf 82}, 74 (1994).

  \bibitem{Peterson}
     C.~Peterson {\it et al.}, Phys.~Rev.~D~{\bf 27} (1983) 105.

  \bibitem{QQ} P. Avery, K. Read, and G. Trahern, Cornell Internal
        Note CSN-22, 1985 (unpublished).

  \bibitem{ref:cft_eff_param}
  This efficiency parameterization followed the functional form
  \[ A \cdot {\rm Freq}\left(  \frac{ \pt - B_1  }{ C_1 } \right)
         \cdot {\rm Freq}\left(  \frac{ \pt - B_2  }{ C_2 } \right)
  \]
  where $A = 0.927$, $B_1 = 6.18$ GeV/$c$, $C_1 = 4.20$,
   $B_2 = 7.48$ GeV/$c$, and $C_2 = 0.504$.


  The efficiency curve turns on at $\pt=6$~\gevc,
  rises to 50\% at $\pt=8$~\gevc,
  and has  a plateau value of 93\%.
  

  \bibitem{Isajet} F.~Paige and S.~D.~Protopopescu, Brookhaven National Laboratory
     Report No.~BNL 38034, 1986.



  \bibitem{PDG} R. M. Barnett {\it et al.}  (The Particle Data Group),
        Physical Review D{\bf 54}, 1 (1996).

  \bibitem{Minuit} F. James, CERN Program library long writeup D506.


\end{thebibliography}
\end{document}